\documentclass[a4paper,11pt]{article}
\usepackage{cite}
\usepackage{amsmath,amsthm}
\numberwithin{equation}{section}
\usepackage{amsfonts}
\usepackage{geometry}
\usepackage{indentfirst}
\usepackage{graphicx}
\usepackage{xcolor}
\usepackage{float}
\usepackage{subfigure}
\usepackage{appendix}
\usepackage{amssymb}
\usepackage{hyperref}
\usepackage{authblk}
\usepackage{fancyhdr}
\usepackage{jheppub}
\usepackage[normalem]{ulem}

\hypersetup{
colorlinks=true,
linkcolor=blue,
filecolor=red,
urlcolor=red,
citecolor=red,
linktocpage=true
}
\newcommand{\be}{\begin{equation}}
\newcommand{\ee}{\end{equation}}

\preprint{PCFT-25-43, LITP-25-01}

\title{
Quantum Corrections in the Low-Temperature Fluid/Gravity Correspondence}
\author[1,2]{Jun Nian}
\author[3]{Leopoldo A. Pando Zayas}
\author[1]{and Cong-Yuan Yue}

\affiliation[1]{International Centre for Theoretical Physics Asia-Pacific,\newline University of Chinese Academy of Sciences, 100190 Beijing, China}
\affiliation[2]{Peng Huanwu Center for Fundamental Theory, Hefei, Anhui 230026, China}
\affiliation[3]{Leinweber Institute for Theoretical Physics, University of Michigan\\
Ann Arbor, MI 48109, USA}

\date{}

\abstract{
Attempts to construct a low-temperature version of the fluid/gravity correspondence have faced obstacles manifested in the form of logarithmic terms in the frequency, $\log(\omega)$, leading to non-local in time constitutive relations for the stress tensor and the charge current. These difficulties can be broadly presented as a breakdown of the hydrodynamic description due to additional infrared modes. We employ new quantum insights into the physics of near-extremal black holes, brought about in the context of Jackiw-Teitelboim gravity, as an effective description of quantum fluctuations in the throat, to revisit the fluid/gravity correspondence at very low temperatures. The quantum corrections naturally include a new length scale, $C$, and an effective action that parametrizes the breaking of near-horizon symmetries. We show that with an appropriate choice of order of limits in the derivative expansion within the low-temperature regime, the $\log(\omega)$ infrared divergence can be resolved. By quantum averaging the infrared Schwarzian modes as an effective extra contribution to the long-wavelength fluid modes, the resulting low-temperature effective fluid description is consistent. We present the dispersion relations for all the relevant hydrodynamic modes. We also revisit the shear viscosity to entropy density ratio and find that at very low temperatures the universal $\frac{1}{4\pi}$ bound is violated due to  quantum correction.}

\begin{document}
\maketitle

\section{Introduction}

Fluid dynamics describes systems near local thermodynamic equilibrium following the effective field theory paradigm \cite{landau2013fluid, Nonequilibrium}. In the regime of late times and long wavelengths, it is natural to assume that a field theory reaches local equilibrium. The microscopic dynamical degrees of freedom of the underlying field theory can be effectively described by a few slowly varying macroscopic fluid quantities such as temperature, $T$, chemical potential, $\mu$, and fluid velocity, $u_{\mu}$. The dynamics of these quantities is then governed by conservation equations for the fluid stress-energy tensor and the charge current.

In the AdS/CFT correspondence \cite{Aharony:1999ti},  which posits an equivalence between certain quantum field theories and gravity in asymptotically AdS spacetimes, the dynamics of the bulk geometry corresponds to the dynamics of the boundary conformal field theory. The equations of motion of the boundary fluid, organized in its derivative expansion, are dual to Einstein and Maxwell equations with negative cosmological constant on the bulk side order by order in a derivative expansion; this relationship is called the fluid/gravity correspondence \cite{Policastro:2002se, Bhattacharyya:2007vjd, Rangamani:2009xk, Bredberg:2011jq, Hubeny:2011hd}. The fluid/gravity correspondence has been extended to a number of situations, including forced fluids, nonlinear fluids, and superfluids \cite{Bhattacharyya:2007vs, Bhattacharyya:2008ji, Bhattacharyya:2008mz, Bhattacharyya:2008kq, Gupta:2008th, Hubeny:2010wp, Bhattacharya:2011eea, Hubeny:2011hd, Magan:2014dwa}. The holographic construction of fluid models has provided some insight into even experimentally relevant situations. Two systems often mentioned in this context are the quark-gluon plasma, a plasma of strongly interacting quarks and gluons produced in relativistic heavy-ion collisions, and ultracold atomic Fermi gases, which are very dilute clouds of atomic gases confined in optical or magnetic traps. Both systems have been shown to exhibit a low shear viscosity to entropy density ratio \cite{Kovtun:2004de} which is characteristic of quantum fluids described by holographic duality \cite{adams2012strongly,Cremonini:2011iq}.

Besides the interest in ultracold fluids, the ubiquity of near-extremal solutions on the gravitational side of the AdS/CFT correspondence makes it interesting to explore the fluid/gravity correspondence at very low temperatures.  Indeed, there have been various relevant discussions \cite{Faulkner:2009wj, Edalati:2009bi, Edalati:2010hk, Edalati:2010pn, Davison:2013bxa, Moitra:2020dal, Arean:2020eus}.  In particular, in \cite{Moitra:2020dal}, the fluid/gravity correspondence was extended to the near-extremal RN-AdS black brane case. In contrast to the high-temperature regime given by $\omega,k_{x}\ll T,\mu$, a different low-temperature regime was considered in \cite{Moitra:2020dal}:
\begin{equation}
   T\ll \omega, k_{x} \ll \mu\, ,
\end{equation}
which interprets the temperature $T$ as the lowest scale of the system,  also previously explored in  \cite{Davison:2013bxa}. Within such a regime, the near-horizon geometry of an AdS$_4$ black brane becomes AdS$_2 \times \mathbb{R}^2 $ given by the metric:
\begin{equation}
    ds^2 =6(r-r_{h})^2dv^2 +2dvdr+r^2_{h}(dx^2+dy^2)\, ,
\end{equation}
which admits a rescaling symmetry
\begin{equation}
    v\to \lambda v\, , \quad (r-r_h)\to \frac{1}{\lambda}(r-r_h)\, .
\end{equation}
This symmetry implies that terms containing the frequency of the perturbation, $\omega$, are always important near the horizon.  This scaling symmetry is ultimately responsible for the appearance of $\log \omega$ in the perturbation, leading to non-local terms in the constitutive relations, as presented in \cite{Moitra:2020dal}, which signals the breakdown of the derivative expansion.

An insightful model-independent framing of the low-temperature hydrodynamic approach was provided in \cite{Arean:2020eus}, which recasts the problem in terms of the diffusive poles in the retarded Green's function in the framework of a near-extremal holographic fluid.  In the low-temperature regime, there is a collision between the diffusive pole of retarded Green's function and a pole associated with the $AdS_2$ throat geometry. Since diffusive poles correspond to hydrodynamic modes, the existence of an $AdS_2$ mode shows that there might exist an extra degree of freedom which is not included in the original effective long-wavelength fluid description. The collision of poles reveals a coupling of hydrodynamic modes with the infrared mode, indicating the breakdown of the hydrodynamic description at very low temperatures.

Recent lessons from the study of two-dimensional Jackiw-Teitelboim gravity \cite{Almheiri:2016fws, Maldacena:2016upp, Jensen:2016pah} show the importance of quantum effects, which are responsible for explicit and spontaneous symmetry breaking in this theory. When applied to higher-dimensional black holes, we are taught that it is perilous to directly take the zero-temperature (extremal)  limit to arrive at an exact AdS$_{2}$ geometry without considering the role of quantum fluctuations, which become strongly coupled as we take the temperature to zero. For near-extremal black holes, when going to very low temperatures, there exists a scale, $T_q$, below which semi-classical thermodynamics breaks down as anticipated in \cite{Preskill:1991tb, Maldacena:1998uz, Page:2000dk}. The dynamics of JT gravity can be described by an emergent boundary Schwarzian theory, and the Euclidean path integral of this theory can be computed exactly \cite{Almheiri:2014cka, Maldacena:2016upp, Stanford:2017thb}. When the temperature goes to the scale of $T\sim T_q$, the $SL(2,\mathbb{R})$ conformal symmetry of AdS$_{2}$ is broken, and the emergent boundary Schwarzian theory becomes strongly coupled. Including the quantum fluctuations of the emergent boundary Schwarzian mode leads to a $\log (T/T_{q})$ correction for the free energy, crucially modifying the low-temperature thermodynamics of near-extremal black holes \cite{Iliesiu:2020qvm}. Similar results have been established in a number of situations \cite{Heydeman:2020hhw, Boruch:2022tno, Moitra:2021uiv, Nanda:2023wne, Turiaci:2023wrh, Kapec:2023ruw, Rakic:2023vhv, Maulik:2024dwq}. {\it Motivated by this state of affairs, we consider the role of low-temperature quantum fluctuations in the fluid/gravity correspondence.}

We consider the effect of quantum fluctuations for the near-extremal 4d Reissner-Nordstrom (RN) black brane and its dual boundary fluid description. Upon dimensional reduction, the effective low-energy theory is described by the Schwarzian mode, which emerges at the boundary of the near-horizon region (NHR) and captures the low-energy gravitational quantum fluctuations.  We integrate out the Schwarzian quantum fluctuations in $Diff(S^1)/SL(2,\mathbb{R})$, which leads to an effective equation of motion with a quantum correction for solving the metric perturbation. This quantum-averaged equation of motion can be naturally interpreted as a Schwinger-Dyson equation: $\langle \frac{\delta S}{\delta \phi} \rangle=0$. Having a mechanism to treat fluctuations enables us to establish the conventional fluid/gravity correspondence and obtain the quantum-corrected boundary fluid stress tensor using the holographic renormalization method.

We will first link the emergence of $\log(\omega)$ terms from the equation of motion to the result of a certain order of limits. The conditions that determine the appropriate order of limits are:

\begin{itemize}
    \item  Scale of perturbations: The energy scale of the external perturbation $\omega$ should always be smaller than the energy scale of this system $T$ (or $T_q$), otherwise a single external perturbation will drastically break the equilibrium state and the long-wavelength description. This is similar to the breakdown of thermodynamics in near-extremal black holes as discussed in \cite{Preskill:1991tb}, where a single Hawking quanta invalidates thermodynamics.
    
    \item  $SL(2,\mathbb{R})$ symmetry breaking: For relatively high temperatures ($T>T_q$), the temperature correction on the near-horizon geometry explicitly breaks the $AdS_2$ conformal symmetry. For very low temperatures ($T<T_q$), by gauge fixing $SL(2,\mathbb{R})$ and averaging over the  Schwarzian mode in the path integral as effective contribution, the time rescaling symmetry (responsible for the $\log\omega$ terms) in the near-horizon region is also fixed, and the transformation cannot be performed again.
\end{itemize}
    
The two points above explain precisely why and how the rescaling symmetry is broken, leading to an expansion that avoids the offending $\log(\omega)$ terms. Previous approaches that take the $T\to 0$ limit first and work directly with the AdS$_2$ metric in the throat region \cite{Edalati:2009bi, Edalati:2010hk, Edalati:2010pn, Moitra:2020dal} neglect the coupling to the Schwarzian mode. In this manuscript, within the context of explicit and spontaneous symmetry breaking developed in the JT framework, we calculate the quantum-corrected stress tensor and charge current by averaging the emergent infrared mode as an effective contribution to the hydrodynamic modes.  We also provide an interpretation in which the very low-energy hydrodynamics is consistently modified by allowing it to couple to an extra infrared mode following a paradigm discussed in \cite{Arean:2020eus}.

The rest of the paper is organized as follows. In Sec.~\ref{sec:Review} we review the gravitational background and key aspects of the fluid/gravity correspondence; we also present the natural appearance of the infrared $\log(\omega)$ terms in the standard approach.  In Sec.~\ref{sec:Quantum Corrections to Equation}, we set up a general method for considering quantum fluctuations at the level of the equation of motion. In Sec.~\ref{sec:Massless Scalar Model}, we use the massless scalar as a prototype to discuss the solution of the quantum-corrected equation of motions and show how appropriately taking the low-temperature limit leads to a resolution to the $\log(\omega)$ problem. In Sec.~\ref{sec:Near-Ext Fluid Gravity Correspondence}, we follow the conventional fluid/gravity correspondence process and calculate the quantum-corrected first-order stress tensor and charge current for boundary fluid, leading to explicit answers for $\eta/s$ as well as various dispersion relations.  In Sec.~\ref{sec:Discussion}, we discuss our results and future prospects in more detail. We relegate some technical aspects to several appendices.

\section{Brief Review of the Fluid/Gravity Correspondence}\label{sec:Review}

There are various excellent reviews in the fluid gravity correspondence to which we direct the reader for more details \cite{Rangamani:2009xk, Hubeny:2011hd}.
In this section, we will provide a brief review of the fluid gravity correspondence and highlight the main issue that arises at very low temperatures.

\subsection{The Gravity Background}

We focus on one of the simplest non-trivial frameworks for the fluid/gravity correspondence, which requires ingredients that generate the equations of conservation for the energy-momentum tensor and a certain current. On the gravitational side, this setup  involves the Einstein-Maxwell action with a negative cosmological constant:
\begin{equation}\label{Eq:Einstein-Maxwell-L}
    I=\frac{1}{16\pi G}\int d^4x \sqrt{-g}\, (R-2\Lambda -F_{MN}F^{MN})\, ,
\end{equation}
where $\Lambda=-3$, and the indices $M,N$ denote the 4d bulk spacetime. We consider the Reissner-Nordstr\"om (RN) black brane solution, which is given by
\begin{equation}
\begin{aligned}
    g_{MN}dx^{M}dx^{N} & = -r^2f(r)dt^2+\frac{dr^2}{r^2f(r)}+r^2(dx^2+dy^2)\, ,\\
      A_{M}dx^{M} & = g(r)dt\, ,
\end{aligned}
\end{equation}
where
\begin{equation}
    \begin{aligned}
        f(r) & = 1-\frac{2GM}{r^3}+\frac{Q^2}{r^4}\, ,\\
        g(r) & = -\frac{Q}{r}\, .
    \end{aligned}
\end{equation}
For a black brane at extremality, the energy density and the charge density are \cite{Moitra:2020dal}
\begin{equation}
    \begin{aligned}
GM=2r_h^3 ,\quad Q=\sqrt{3}r_h^2\, ,\label{Mass charge in extremality}
    \end{aligned}
\end{equation}
where $r_h$ is the extremal horizon location, and the extremal chemical potential is given by $\mu= \sqrt{3}r_h$, and temperature is defined as $T=\frac{3r_+^4 -Q^2}{4\pi r_{+}^3}$ with outer horizon location and charge. The near-extremal expressions for black hole mass and entropy are
\begin{equation}
    \begin{aligned}
        M&=\frac{2r_h^3}{G}+\frac{9r_h}{4G}T^2+\mathcal{O}(T^3)\\
        S&=\frac{A}{4G}=\frac{\pi r_{+}^2}{G}=\frac{\pi r_h^2}{G}+\frac{\pi r_h T}{G}+\mathcal{O}(T^2).
    \end{aligned}
\end{equation}
Near extremality, the mass of the black hole grows quadratically and the entropy grows linearly in temperature, which implies that there exists a regime where the black hole energy is too low to emit even a single Hawking quanta. This breakdown of thermodynamics was anticipated in \cite{Preskill:1991tb}. By considering the contribution of the quantum fluctuations in the near-horizon region following the $JT$ model, the authors of \cite{Iliesiu:2020qvm} showed that the energy and entropy receive quantum corrections that ameliorate the breakdown of thermodynamics.

At extremality, the outer horizon location coincides at $r_h$, and we have
\begin{equation}
    GM=2\, r_h^3\, ,\quad Q=\sqrt{3}\, r_h^2\, ,\quad \mu=\sqrt{3}\, r_h\, .
\end{equation}
Now, the extremal AdS$_4$ black brane metric becomes
\begin{equation}
    \begin{aligned}
       f_{0}(r) & =1-\frac{4r_{h}^3}{r^3}+\frac{3r_{h}^4}{r^4} =\frac{(r-r_{h})^2(r^2+2rr_{h}+3r_{h}^2)}{r^4}\, ,
       \\g_{0}(r) & = -\frac{\sqrt{3}r_{h}^2}{r}\label{extremla metric}\, .
    \end{aligned}
\end{equation}
In the near-horizon region $r-r_h \ll r_h$, we can expand the metric with $\frac{r-r_h}{r_h}$. Further using the ingoing Eddington-Finkelstein coordinates
\begin{equation}
    v=t+r^*,\quad r^*=\int \frac{dr}{r^2f(r)}\, ,
\end{equation}
the metric and the gauge field can be written as
\begin{equation}
    \begin{aligned}
       ds^2 & = -r^2f(r)dt^2+\frac{dr^2}{r^2f(r)}+r^2(dx^2+dy^2) 
       \\& \approx -6(r-r_h)^2dv^2 +2dvdr+r^2(dx^2+dy^2)\, ,
       \\A & = A_{M}dx^{M}=g(r)dv\, .
    \end{aligned}
\end{equation}
The near-horizon region (NHR) becomes an AdS$_2 \times \mathbb{R}^2 $ geometry with a rescaling symmetry
\begin{equation}
    v\to \lambda v\, ,\quad (r-r_h)\to \frac{1}{\lambda}(r-r_h)\, .
\end{equation}
For EoM in NHR, the scaling symmetry makes the $\omega$-dependent terms become important and cannot be ignored, which subsequently leads to the non-local $\log(\omega)$ problem. In the next section, we will explain how insights regarding the breaking of this scaling symmetry address the appearance of the $\log(\omega)$ terms.

\subsection{Fluid Equations from Gravity}

The long-wavelength limit of the AdS/CFT correspondence provides a fluid dynamical description of strongly coupled conformal field theories  \cite{Bhattacharyya:2007vjd, Rangamani:2009xk, Bredberg:2011jq, Hubeny:2011hd}. The  Einstein equation in the bulk is dual to the boundary fluid equation of motion, which is the relativistic Navier-Stokes equation, i.e.,
    \begin{equation}
    \begin{aligned}
        R_{MN}-\frac{1}{2}g_{MN}R+\Lambda g_{MN}&=T_{MN}\quad&\Longleftrightarrow&\quad &\partial_{\mu}T^{\mu\nu}=0,\,\\
        \nabla^{M}F_{MN}&=0 \quad &\Longleftrightarrow& \quad &\partial_{\mu}J^{\mu}=0.
    \end{aligned}
    \end{equation}

At the conformal boundary with ($r\to \infty$)  we have the boundary metric
\begin{equation}
    \begin{aligned}
        ds_{bdy}^2  = r^2 (-dv^2 + dx^2 + dy^2)\, ,
    \end{aligned}
\end{equation}
which is conformally flat. 

The central tenet of the gradient expansion is that it enables local equilibrium, allowing fluid quantities to vary slowly with position.  To implement that approach, one first needs to explicitly introduce those quantities in the gravity solution. In the boundary coordinates, by performing a Lorentz boost in the $x^i$ direction with the velocity $\beta_{i}$, we obtain 
\begin{equation}
     \begin{aligned}
        dv' & = \gamma dv - \gamma \beta_{i} dx^i\, ,
        \\ {dx^i}' & = -\gamma \beta_{i} dv + \gamma dx^i\, ,
    \end{aligned}
\end{equation}
where $\gamma=\frac{1}{\sqrt{1-\beta^2}}$. One can now define the  fluid velocity as $u_{\mu}=\gamma(-1,\beta_{i})$, we can rewrite $dv'=-u_{\mu}dx^\mu$ and using $\eta_{\mu\nu}=diag(-1,1,1)$ and $P_{\mu\nu}=\eta_{\mu\nu} + u_{\mu}u_{\nu}$, we can rewrite the background metric and the gauge field as:
\begin{equation}
    \begin{aligned}
           \bar{g}_{MN}dx^{M}dx^{N} & = -r^2f(r;\mu,T)u_{\mu}u_{\nu}dx^\mu dx^{\nu}-2u_{\mu}dx^{\mu}dr + r^2 P_{\mu\nu}dx^\mu dx^{\nu}\, ,
    \\ \bar{A} & = -g(r;\mu,T)u_{\mu}dx^{\mu}\, .
    \end{aligned}
 \end{equation}
 For the perturbation theory, we first promote the fluid quantities to depend on the boundary coordinates and then expand the metric and the gauge field in a power series of  $\frac{\omega}{\mu},\frac{k_x}{\mu}\sim \epsilon$ as
 \begin{equation}
    \begin{aligned}
        g_{MN} & =\sum_{n=0}\epsilon^n g_{MN}^{(n)}(\mu(\epsilon x^\sigma),\, T(\epsilon x^\sigma),\, u_{\mu}(\epsilon x^{\sigma}))\, ,
        \\A_{M} & = \sum_{n=0} \epsilon^n A_{M}^{(n)}(\mu(\epsilon x^\sigma),\, T(\epsilon x^\sigma),\, u_{\mu}(\epsilon x^{\sigma}))\, .
    \end{aligned}
\end{equation}
Then, the zeroth-order perturbation can be directly obtained as 
\begin{equation}\label{quantum-corrected background metric}
    \begin{aligned}
           g_{MN}^{(0)}(x^\sigma)dx^{M}dx^{N} & = -r^2 f(r;\mu(x^\sigma),T(x^\sigma))  u_{\mu} u_{\nu} dx^{\mu} dx^{\nu} - 2 u_{\mu} (x^{\sigma}) dx^{\mu} dr\\
    {} & \quad + r^2P_{\mu\nu}(x^{\sigma})dx^{\mu}dx^{\nu}\, , \\
    A_{M}^{(0)}dx^{M} & = -g(r;\mu(x^{\sigma}),T(x^{\sigma}))\, u_{\mu} (x^\sigma)\, dx^{\mu}\, .
    \end{aligned}
\end{equation}
We also work in the same gauge as the conventional fluid/gravity correspondence:
\begin{equation}\label{Gauge choice}
     g_{rr} = 0 = A_{r}\, ,
 \qquad g_{r\mu} \propto u_{\mu}\, ,
\qquad g^{(0)MN} g_{MN}^{(n)} = 0,\quad n\ge 1\, .
\end{equation}
One then proceeds to solve the perturbation equation order by order systematically.

A boosted black brane on the bulk side leads to the ideal fluid constitutive relation on the boundary side. The derivative of the fluid dynamics quantities, temperature, fluid velocity, and chemical potential, i.e., ($T$, $u_{\mu}$, $\mu$), can produce an $\mathcal{O}(\epsilon)$ quantity. We can use the powers of $\epsilon$ to organize this derivative expansion:
    \begin{equation}
        g=g^{(0)}(T,u_{\mu},\mu)+\epsilon g^{(1)}(T,u_{\mu},\mu)+\epsilon^2 g^{(2)}(T,u_{\mu},\mu)+\cdots
    \end{equation}
dual to 
    \begin{equation}
       T_{\mu\nu}=T_{\mu\nu}^{(0)}+\epsilon T_{\mu\nu}^{(1)}+\epsilon^2 T_{\mu\nu}^{(2)}\cdots
    \end{equation}
    order by order. \\

Including small metric perturbations around the black brane solution allows us to treat them as fields on the original metric background. The perturbative expansion is organized around the original metric background, which is schematically of the form
    \begin{equation}
        \mathbb{D}(g_{MN}^{(n)},A_{M}^{(n)})=s^{(n)}(g_{MN}^{(n-1)}....,A_{M}^{n-1}....).
    \end{equation}
After obtaining those metric perturbation corrections, we can use the holographic renormalization method to calculate the stress tensor for the boundary fluid \cite{Skenderis:2002wp, Balasubramanian:1999re}. The expression for the stress tensor is
\begin{equation}
    8\pi G{T^{\mu}}_{\nu}=-\lim_{r\to \infty}r^{3} \left(K^{\mu}_{\nu}-(K-2){\delta^{\mu}}_{\nu}-({R^{(3)\mu}}_{\nu}-\frac{1}{2}R^{(3)}{\delta^{\mu}}_{\nu})\right)\, ,
\end{equation}
where $K_{\mu\nu}$ is the extrinsic curvature tensor, and the Ricci tensor ${R^{(3)\mu}}_{\nu}$ is constructed on the boundary with the induced metric $\gamma_{\mu\nu}$. The metric on the boundary manifold is 
\begin{equation}
    \widetilde{\gamma}_{\mu\nu}=\lim_{r\to \infty}\frac{1}{r^2}\gamma_{\mu\nu}\, .
\end{equation}
Correspondingly, the boundary gauge field is
\begin{equation}
    \widetilde{A}_{\mu}=\lim_{r \to \infty}A_{\mu}\, ,
\end{equation}
the boundary electromagnetic field strength is 
\begin{equation}
    \widetilde{F}_{\mu\nu}=\partial_{\mu}\widetilde{A}_{\nu}-\partial_{\nu}\widetilde{A}_{\mu}\, ,
\end{equation}
and the boundary charge current is
\begin{equation}
    4\pi G J^{\nu}=\lim_{r\to\infty}\sqrt{-g}F^{\nu r}\, .
\end{equation}
With this process, the stress tensor and the charge current for the boundary fluid can be obtained.

\subsection{Review of Extremal Calculation without Quantum Corrections}

In this subsection, we will review the extremal solution to the classical EoM without quantum correction based on \cite{Moitra:2020dal} with the standard choice of regime
\begin{equation}
    T\ll \omega,k_{x} \ll \mu\, ,
\end{equation}
where the temperature has been set to zero first. Originally, a massless scalar field on the black hole background satisfies
\begin{equation}
    \nabla^{2}\phi=\frac{1}{\sqrt{-g}}\partial_{M}(\sqrt{-g}g^{MN}\partial_{N}\phi)=0\, .
\end{equation}
A constant value $\phi=\phi^{(0)}$ is one of the solutions to this equation. In the framework of the fluid/gravity correspondence, we can take $\phi^{(0)}$ as the zeroth-order perturbation, and we can further make it dependent on the boundary coordinates $\phi^{(0)}=\phi^{(0)}(\epsilon x^{\mu})$. In the Eddington-Finkelstein coordinates, we take the following Ansatz:
\begin{equation}
    \phi=\phi_{c}(r)e^{-i\omega v +ik_{x} x}=\sum_{n=0}\phi^{(n)}(r)e^{-i\omega v +ik_{x} x}\, ,
\end{equation}
where we assume that $\frac{\omega}{\mu},\frac{k_{x}}{\mu}\sim \epsilon\ll 1$. For simplicity, we consider momentum along $k_x$, but restoring the dependence on $k_y$ is immediate due to symmetries. The fluid derivative expansion is carried by $\epsilon$, and the label $n$ in $\phi^{(n)}$ denotes the order of the perturbation solution. The $n$-th order is sourced by the $(n-1)$-th order in the equation. The AdS$_{4}$ boundary condition is
\begin{equation}
    \phi \to \phi^{(0)}(\epsilon x^{\mu}) ,\quad \phi^{(n)}\to 0, \quad \text{when}\, r\to \infty\, .
\end{equation}
Since the $\mathbb{T}^2$ in the NHR AdS$_2\times \mathbb{T}^2$ is compact, we can choose $x\in [0,L_x]$ and $y\in [0,L_y]$. Consequently, the periodic boundary condition for $\phi_m$ in the $x$-direction is
\begin{equation}
    \phi(t,r,x,y) = \phi(t,r,x+L_x,y)=\phi(t,r,x,y+L_y)\,.
\end{equation}
Therefore, we can have
\begin{equation}
    k_x=\frac{2\pi n_x}{L_x},\quad k_y=\frac{2\pi n_y}{L_y},\quad n_x, n_y\in \mathbb{Z}\, .
\end{equation}
Then, the classical equation of motion is 
\begin{equation}
    \frac{d}{dr} \left(r^4 f(r) \frac{d\phi_{c}}{dr}\right) - 2i\omega r \frac{d(r \phi_{c})}{dr} - k_{x}^2 \phi_{c} = 0\, .
\end{equation}
In the first-order perturbation equation, we have
\begin{equation}
    \frac{d}{dr} \left(r^4 f(r) \frac{d\phi^{(1)}}{dr}\right) - 2i\omega r \frac{d(r \phi^{(1)})}{dr} - k_{x}^2 \phi^{(1)} = (2i\omega r +k_x^2)\phi^{(0)}\, .
\end{equation}
Usually, since $\frac{\omega}{\mu},\frac{k_{x}}{\mu}\sim \epsilon\ll 1$ are small quantities, in the first order derivative expansion, we should ignore all ($\omega, k_x$)-dependent terms in the left hand side and $k_x^2$ term in the right hand side. However, here, since the extremal NHR $AdS_2 \times T^2$ geometry has a time rescaling symmetry
\begin{equation}
\begin{aligned}
      ds^2&=-6(r-r_h)^2dv^2 +2dvdr +r_h^2 (dx^2 +dy^2)\\
       &v\to \lambda v\, ,\quad (r-r_h)\to \frac{1}{\lambda}(r-r_h)\, ,
\end{aligned}
\end{equation}
the $\omega$-dependent terms in EoM will always be important in NHR and cannot be ignored. Therefore, we need to divide the spacetime into the NHR and the FAR, which overlap at $r_{B}$, with the condition
\begin{equation}
    \omega\ll r_{B}-r_{h}, \quad r_{B}-r_{h}\ll r_{h}\, .
\end{equation}
We can solve the EoMs in NHR and the FAR separately and then match the solutions at the overlap region to obtain the full solution in the bulk.

\subsubsection{Extremal Solution in the Far-Away Region}

In the FAR, by ignoring the second-order terms in $\frac{\omega}{\mu},\frac{k_{x}}{\mu}\sim \epsilon$, we have a simplified EoM:
\begin{equation}
    \frac{d}{dr} \left(r^2f(r)_{FAR}\frac{d\phi^{(1)}_{out}}{dr}\right) = 2i\omega r \phi^{(0)}\, ,
\end{equation}
where  $f(r)_{FAR}=r^2(1-\frac{4r_h^3}{r^3}+\frac{3r_h^4}{r^4})$ is the extremal FAR metric factor. The solution to this differential equation can be directly expressed as 
\begin{equation}\label{eq:FAR solution}
    \phi^{(1)}_{out}=B^{(1)}_{out}-\int^{\infty}_{r}\frac{dr'}{r'^2 f(r')_{FAR}} \left(A^{(1)}_{out}+\int^{r'}_{r_{B}}dr''2i\omega r''\phi^{(0)}\right)\, .
\end{equation}
Using the boundary condition, $\phi^{(n)}\to 0$ for $r\to \infty$, we obtain $B_{out}^{(1)}=0$ and
\begin{equation}
    \int^{r'}_{r_{B}}dr'' 2i\omega r'' \phi^{(0)}=i\omega \phi^{(0)}(r'^2-r_{B}^2)\, .
\end{equation}
The integral in \eqref{eq:FAR solution} can be directly evaluated:
\begin{equation}
    \begin{aligned}
    &\int^{\infty}_{r}\frac{dr'}{r'^2F(r')}\left (A^{(1)}_{out}+\int^{r'}_{r_{B}}dr''2i\omega r''\phi^{(0)}\right )\\
    =\,\, & - \Bigg[-\frac{1}{6r_{h}^2}\frac{A_{out}^{(1)}-i\omega \phi^{(0)}(r_{B}^2+2r_{h}^2)}{r-r_{h}}-\frac{A_{out}^{(1)}-i\omega\phi^{(0)}(r_{B}^2+2r_{h}^2)}{18r_{h}^3}\log\frac{(r-r_{h})^2}{r^2+2rr_{h}+3r_{h}^2}
    \\ & \qquad - \frac{A_{out}^{(1)}-i\omega \phi^{(0)}(r_{B}^2-7r_{h}^2)}{36\sqrt{2}r_{h}^3} \bigg(\pi -2\arctan \Big(\frac{r+r_{h}}{\sqrt{2}r_{h}} \Big) \bigg)\Bigg]\, .
    \end{aligned}
\end{equation}
Near $r=r_{B}$, we can do an expansion with respect to $\frac{r-r_{h}}{r_{h}}$:
\begin{equation}\label{eq:Expand phi_out}
    \phi_{out}^{(1)}=-\frac{1}{6r_{h}^2}\frac{A_{out}^{(1)}-i\omega\phi^{(0)}( r_{B}^2 - r_{h}^2)}{r-r_{h}}-\frac{A_{out}^{(1)}-i\omega\phi^{(0)}(r_{B}^2+2r_{h}^2)}{9r_{h}^3}\log\frac{r-r_{h}}{r_{h}}+\mathcal{O} \left[\left(\frac{r-r_{h}}{r_{h}} \right)^{0}\right]\, ,
\end{equation}
where the constant will be determined by matching with the NHR solution.

\subsubsection{Extremal Solution in the Near-Horizon Region}

The metric in the NHR is
\begin{equation}
    ds^2 = -f(r)_{NHR}dv^2 + 2dvdr + r_h^2 (dx^2 + dy^2)\, ,
\end{equation}
where $f(r)_{NHR}=6(r-r_{h})^2$. We can neglect the terms $\sim k_{x}^2$ in the EoM, but since the metric has a time rescaling symmetry, we can always rescale $v$ to make the term $\sim \omega$ in the EoM important \cite{Moitra:2020dal}. Then, the classical EoM for the first-order perturbation can be written as 
\begin{equation}
    \frac{d}{dr} \left(f(r)_{NHR}\frac{d\phi_{in}^{(1)}}{dr}\right) - 2i\omega \frac{d\phi_{in}^{(1)}}{dr}=\frac{2i\omega \phi^{(0)}}{r_{h}}\, .
\end{equation}
By defining $r^{*}=\int \frac{dr}{f(r)_{NHR}}$, we can rewrite the EoM as
\begin{equation}
    \frac{d^2\phi^{(1)}_{in}}{dr^{*2}} - 2i\omega\frac{d\phi^{(1)}_{in}}{dr^{*}} = \frac{2i\omega\phi^{(0)}}{r_{h}}F(r)\, .
\end{equation}
Then, the interior solution can be generally rewritten as
\begin{equation}
    \phi^{(1)}_{in} = A^{(1)}_{in}+B^{(1)}_{in}e^{2i\omega r^{*}} + \frac{2i\omega\phi^{(0)}}{r_{h}}e^{2i\omega r^{*}}\int^{r^{*}}_{-\infty}dr^{*'}e^{-2i\omega r^{*'}}\int^{r'}_{r_{h}}dr^{''}\, ,
\end{equation}
where
\begin{equation}
    r^{*}=-\frac{1}{6(r-r_h)}\,,
\end{equation}
Imposing an ingoing boundary condition on the horizon can set $B_{in}^{(1)}=0$. Then we have
\begin{equation}
\begin{aligned}
        \phi_{in}^{(1)}&=A_{in}^{(1)}+\frac{i\omega \phi^{(0)}}{3r_h}e^{2i \omega r^{*}}\int_{-2\zeta}^{\infty}\frac{dt}{t}e^{it}\,\\
        &=A_{in}^{(1)}+\frac{i\omega \phi^{(0)}}{3r_h}e^{2i \omega r^{*}}E_1 (2i \zeta)\, ,
\end{aligned}
\end{equation}
where $\zeta =\omega r^{*}=-\frac{\omega}{6(r-r_h)}$ and $E_1(2i\zeta)$ is the exponential integral function. Near the boundary of NHR (overlap region), we have $|\omega r^{*}|\ll 1$, the function can be expanded as
\begin{equation}
\begin{aligned}
        E_1 (2 i\omega r^{*})&=-\gamma -\log(2i \omega r^{*})+\mathcal{O}(\omega r^{*})\,\\
        &=\frac{i\pi}{2} -\gamma +\log 3 + \log \frac{r-r_h}{\omega} +\mathcal{O} \left(\frac{\omega}{r-r_h} \right)\, ,
\end{aligned}
\end{equation}
where $\gamma\approx 0.5772$ is the Euler-Mascheroni constant. The asymptotic behavior of the inner solution near the NHR boundary is
\begin{equation}
    \phi^{(1)}_{in} = A_{in}^{(1)} + \frac{i\omega \phi^{(0)}}{3r_h} \left(\frac{i\pi}{2} - \gamma + \log 3 \right) + \frac{i\omega \phi^{(0)}}{3r_h} \log\left(\frac{r-r_h}{\omega}\right) + \cdots\,.
\end{equation}
Matching $\phi^{(1)}_{out}$ and $\phi^{(1)}_{in}$ at the overlap region, one can find that since the inner solution do not have $\frac{1}{r-r_h}$ term, based on \eqref{eq:Expand phi_out}, $A_{out}^{(1)}$ in the FAR solution should be given by
\begin{equation}
    A_{out}^{(1)}=i\omega \phi^{(0)}(r_{B}^2-r_{h}^2)\, .
\end{equation}
We can also fix the $A_{in}^{(1)}$ by matching the coefficient as
\begin{equation}
    A_{in}^{(1)} = \frac{i\omega \phi^{(0)}}{3r_h}\log \frac{\omega}{r_h} - \frac{i\omega \phi^{(0)}}{12r_h} \left((\sqrt{2}+2i)\pi + 6\log 3 + 2\log 2 -2\sqrt{2}\tan^{-1}\sqrt{2} - 4\gamma \right)\,.
\end{equation}
Thus, both inner and outer solutions have been fixed. As we can see, there is a $\log (\omega)$ term, which signals the breakdown of fluid derivative expansion.

\textbf{\flushleft About the non-analytic $\log(\omega)$ term:}

In the extremal discussion, we have seen that the time rescaling symmetry can enhance the $\omega$-dependent term in the EoM, and subsequently lead to the emergence of a logarithmic term. Also, in the near-extremal discussion of \cite{Moitra:2020dal} (see also  \cite{Gouteraux:2025kta} for a more recent discussion), it have shown that the inner solution contains a polygamma function $\psi(1-\frac{i\omega}{3T})$, which also gives $\log \omega$ term when considering the $\frac{\omega}{T}\gg 1$ limit. The physical reason behind it could be that the choice of the expansion parameter is no longer appropriate due to the near-horizon AdS$_{2}$ rescaling symmetry, or the temperature regime choice is improper for this case.

From a mathematical perspective, the reason why the non-analytic term appears is from the asymptotic behavior of the digamma function\cite{handbook}:
\begin{equation}
    \textrm{Re}\, \psi(1-i y)=\log y + \frac{1}{12 y^2} + \frac{1}{120y^4} + \frac{1}{252 y^6} + \cdots \quad\text{for}\quad y\to \infty\, .
\end{equation}
On the contrary, when $y$ is very small, the $\log(y)$ term disappears in the asymptotic behavior of the digamma function:
\begin{equation}
    \psi(1-iy)=-\gamma -\frac{1}{6}i\pi^2 y + \mathcal{O} (y^2)\, .
\end{equation}
At the EoM level, one should consider that all ($\omega, k_x$)-dependent terms disappear on the l.h.s of 
\begin{equation}
    \frac{d}{dr} \left(f(r)_{NHR}\frac{d\phi_{in}^{(1)}}{dr}\right) - 2i\omega \frac{d\phi_{in}^{(1)}}{dr}=\frac{2i\omega \phi^{(0)}}{r_{h}}\, ,
\end{equation}
to make sure the perturbation equation is well defined in each perturbation level n. In this sense, to eliminate the $\log\omega$ term is to eliminate the $\omega$-dependent term in the l.h.s of EoM.

Therefore, two kinds of possibilities can make the $\log(\omega)$ term disappear in the low-temperature fluid/gravity correspondence.
\begin{itemize}
    \item[a)] By choosing a different regime:
    
    Within the regime of $\omega\ll T\ll \mu$, we have $\frac{\omega}{T}\ll 1$, which corresponds to a relatively high temperature case, in which the NHR rescaling symmetry is explicitly broken. Therefore, $\psi(1-i\frac{\omega}{3T})$ will no longer contribute a $\log(\omega)$, and in the EoM level, the $\omega$-dependent term can be ignored at first place.

    \item[b)] By quantum averaging the Schwarzian mode.

    First of all, the quantum average of the Schwarzian mode will give an extra energy scale $\frac{1}{C}$, which gives new possible regime choices. Secondly, the quantum average Schwarzian mode requires gauge fixing of $SL(2,\mathbb{R})$ gauge redundancy in the path integral, which ensures that in the very low-temperature case, one can no longer perform the time rescaling transformation and make the $\omega$-dependent term important. 
\end{itemize}
Here is a brief sketch of the basic idea of removing the non-analytic term. We will make a more detailed discussion about this term in Sec.~\ref{sec: non-analytic term}. Before moving on, it is worth making some important remarks on the extremal treatment:

Firstly, we should note that it is inappropriate to directly extend the semi-classical analysis of the conventional fluid/gravity correspondence to near-extremal or extremal cases, since semi-classical thermodynamics is no longer valid in this regime. When one needs to extend the semi-classical discussion to the near-extremal case, considering quantum corrections (especially considering the contribution of infrared zero modes) is necessary. With the consideration of quantum correction, the entropy of a near-extremal black hole has $-\frac{3}{2}\log T$ correction, which will lead the entropy to minus when naively taking a $T\to 0$ limit. Therefore, to ensure we always remain in a regime where thermodynamics is valid, we should not take the temperature strictly to zero, but rather stay at a very low, finite value.

Secondly, the appearance of $\log\omega$ in this case is due to the perfect $AdS_2$ time rescaling symmetry at extremality, which can always enhance the importance of the $\omega$-dependent term in the EoM. Later, we will show explicitly via a discussion on a polygamma function appearing in the solution at the EoM level that, for some appropriate regime choice, which preserves the local equilibrium of the fluid, the $\log(\omega)$ term disappears and the non-analytic problem in the derivative expansion can be resolved.

\section{Quantum Corrections and Equations of Motion}\label{sec:Quantum Corrections to Equation}
The physics of the throat region is captured by a version of the Jackiw-Teitelboim theory. This sector describes low-temperature physics. The immediate implications for the thermodynamics have been established as modifications of the path integral; these contributions dominate over the semiclassical result \cite{Iliesiu:2020qvm}.  In this manuscript, we argue that similar path-integral corrections are central to the fluid/gravity correspondence.

The standard method for incorporating quantum aspects into the gravitational setup takes the general form of averaged quantum fields in a fixed classical background, $G_{\mu\nu}=\kappa \langle T_{\mu\nu}\rangle$. The framework is that fields can be treated quantum-mechanically in a fixed gravitational background. Here, we need to push beyond this paradigm as the metric itself is not classical. In this section, we explain the approximation we use, which includes quantum fluctuations in the metric. The main result is Eq.~\eqref{Eq:modified metric}.

\subsection{Quantum-Corrected Thermodynamics}

In this subsection, we review the quantum-corrected RN black brane thermodynamics from the holographic point of view, with an emphasis on the holographic renormalization framework. We will demonstrate how a modification to the classical metric, followed by the application of holographic renormalization, yields results that are consistent with the more rigorous partition function approach to thermodynamics.

To illustrate the relevance of JT gravity, we start from the action \eqref{Eq:Einstein-Maxwell-L}, and perform a dimensional reduction along the directions $(x,y)$ which form a torus, $\mathbb{T}^2$.  We obtain a 2d dilaton gravity model whose leading sector is JT gravity \cite{Iliesiu:2020qvm}. The action takes the following form:\footnote{Because we are considering a black brane with planar symmetry, we note slight differences between our reduction result above compared to  Eq.~(2.26) in \cite{Iliesiu:2020qvm}. By dimensional reduction of the 4d action to 2d and integrate the 2d $U(1)$ gauge field we will obtain a 2d dilaton gravity model with different dilaton potential $U(\chi)=r_{h}(\frac{Q^2}{ \chi^{\frac{3}{2}}}-\frac{3\chi^{\frac{1}{2}}}{L^2})$, where $\chi(r) $ is the original nonlinear 2d dilaton field which is further linearly expanded to $\frac{\chi(r)}{G}=\Phi_{0}+\Phi(r)$ in the near horizon region. Also, the volume of the compactified space is ${\rm vol}(T^2) =L_x L_y$. We will set $L_x L_y = 4\pi$ for simplicity in later discussions.}
\begin{equation}
    \begin{aligned}
        I_{EM}[g_{\mu\nu}^{(2d)},\chi] & = \beta M_{0}-\frac{1}{4}\int d^2x \sqrt{g}\, \left[\Phi_{0} R + \Phi \left(R + \frac{2}{L_{2}^2}\right) + \mathcal{O}\left(\frac{\Phi^2}{\Phi_{0}^2}\right)\right]\\
        & \quad -\frac{1}{2}\int du \sqrt{h}\, \left[\Phi_{0}K_{NHR}+\frac{\Phi_{b}}{\epsilon}\left(K_{NHR}-\frac{1}{L_{2}}\right)\right]\, ,\label{reducedaction}
    \end{aligned}
\end{equation}
where  $\Phi$ is the 2d dilaton field and $\Phi_{0}$ is the dilaton value at the horizon $r_h$, $L_2$ is the $AdS_2$ radius and $\epsilon$ is the small cut-off at the $AdS_2$ boundary and after evaluating the extrinsic term the $\epsilon$ can be canceled and only remains in the Schwarzian derivative.

The standard way of computing the thermodynamics of this action is to perform the path integral of the JT/Schwarzian action and to  subsequently use the partition function $Z_{RN}(\beta,\phi_b)$ to calculate the thermodynamic quantities. The exact partition function is given by 
\begin{equation}\label{Eq:JT-Partition-Function}
    Z_{RN} = \left(\frac{\Phi_{b}}{\beta}\right)^{3/2}\, e^{\pi \Phi_0-\beta M_0 +\frac{2\pi^2}{\beta}\Phi_b}\, .
\end{equation}
The energy and entropy can be calculated as
\begin{equation}
    \begin{aligned}
        E&=-\partial_{\beta}\log Z_{RN}=M_0 +\frac{2\pi^2\Phi_b}{\beta}+\frac{3}{2\beta},\\
        S&=(1-\beta \partial_\beta)\log Z_{RN}=S_0+\frac{4\pi^2 \Phi_b}{\beta}-\frac{3}{2}\log(\frac{\beta}{\Phi_b}).
    \end{aligned}
\end{equation}
These results were first presented in \cite{Iliesiu:2020qvm}, where $\Phi_b=C=T_{q}^{-1}=\frac{r_h L_2^2}{G}=\frac{r_h }{6G}$. The thermodynamics is modified in two crucial ways: there is a correction to the energy linear in $T$ and a correction to the entropy logarithmic in $T$. These two modifications alleviate the thermodynamic instability of near-extremal black holes first articulated in \cite{Preskill:1991tb}. 

Given that in the fluid/gravity correspondence we are concerned with the dynamical perturbative behavior in the bulk gravity, it becomes a natural starting point to reproduce the above quantum-corrected thermodynamics using the holographic framework.

Consider the extremal NHR metric in ingoing Eddington-Finkelstein coordinates
\begin{equation}
    ds^2 =-6(r-r_h)^2 dv^2 +2dvdr+r_h^2 (dx^2 + dy^2) \, .
\end{equation}
This is a pure $AdS_2 \times \mathbb{T}^2$ geometry. In the NHR $AdS_2$ boundary ($\gamma =-dv^2$), any reparameterization of time $u=u(v)$ can be compensated by a Weyl rescaling $\gamma_{\mu\nu}\to e^{2\Omega}\gamma_{\mu\nu}$ and leaves the metric invariant. On the bulk side, these are conformal transformation corresponding to diffeomorphisms which preserve the radial gauge and fix the boundary metric\footnote{Here the physical mode are large diffeomorphisms $Diff(S^1)/SL(2,\mathbb{R})$, and small diffeomorphisms $SL(2,\mathbb{R})$ are residual gauge degrees of freedom, can be gauge fixed when performing quantization\cite{Moitra:2021uiv}.}. Under boundary time reparametrizations $u(v)$, the previous $AdS_2$ vacuum geometry becomes \cite{Jensen:2016pah} 
\begin{equation}
     ds^2 = -6 \left((r-r_h)^2 + 2 Sch(u(v),v)\right) dv^2 + 2 dv dr + r_h^2 (dx^2 + dy^2),
\end{equation}
where $Sch(u(v),v)$ is the Schwarzian derivative. A saddle point of the transformation $u(v)=\tanh(\pi T v)$ yields 
\begin{equation}\label{classical Schwarzian derivative}
    Sch(u(v),v)=-2\pi^2 T^2,
\end{equation}
which corresponds to an $AdS_2$ black hole metric
\begin{equation}
    ds^2 = -6 \left((r-r_h)^2 -4\pi^2 T^2\right) dv^2 + 2 dv dr + r_h^2 (dx^2 + dy^2)\, ,
\end{equation}
with horizon location $r_{\pm}=r_h \pm 2\pi T$. This demonstrates that the Schwarzian mode can indeed capture the expected low-temperature excitations away from the extremality of the black hole. Holographic renormalization of this metric was implemented in  \cite{Jensen:2016pah}, the stress tensor at the $AdS_2$ boundary was shown to be
\begin{equation}
    \langle T^{\mu\nu}\rangle=\frac{2}{\sqrt{h}}\frac{\delta S_{ren}}{\delta \gamma_{\mu\nu}}=\Phi_b \gamma^{\mu\nu}Sch(u(v),v).
\end{equation}
Our primary technical challenge is to extend this stress-energy tensor to one corresponding to the asymptotic $AdS_4$ boundary. This procedure requires matching the asymptotic behavior of the NHR metric and the FAR metric at the overlap region and imposing continuity to glue on the NHR quantum correction factor to the FAR. This can be interpreted as extending the RG flow from IR to AdS$_4$ boundary at $r\to \infty$.\footnote{Note that the continuity condition in the overlap region might not fix the geometry uniquely; one must further require that the RG flow to $AdS_4$ should reproduce the same thermodynamics consistent with \cite{Iliesiu:2020qvm}.}
Within these considerations, the FAR metric is given by
\begin{equation}
    ds^2 = -\frac{\left((r-r_h)^2 + 2 Sch(u(v),v)\right)\, (r^2 +2rr_h +3r_h^2)}{r^2}dv^2 +2dvdr +r^2 (dx^2 +dy^2)\, .
\end{equation}
We can average the background metric in the Schwarzian path integral and denote the quantum-corrected background as
\begin{equation}\label{Eqn:BackgroundMetric}
\begin{aligned}
        &ds^2 =-\frac{(r-r_h)^2(r^2 +2rr_h +3r_h^2)}{r^2}Q_{cor}(r)dv^2 +2dvdr +r^2 (dx^2 +dy^2)\, \\
        &Q_{cor}(r)=1+\frac{2\langle Sch(u(v),v)\rangle}{(r-r_h)^2}\,.
\end{aligned}
\end{equation}
Let us address several issues that the approach of correcting the metric raises. First, it is customary to use the Fefferman-Graham expansion of the metric to read off the boundary stress-energy tensor via holographic renormalization. In standard holographic renormalization, the metric in the bulk is generated as an expansion of the solution to the Einstein equation, that is, it is generated as an on-shell solution. Since the diffeomorphisms introduced in the AdS$_2$ region are not a symmetry of AdS$_4$, we need to contend with some amount of off-shellness. However, it is possible to rewrite this extra correction factor back into action as an effective action and perform the holographic renormalization in an on-shell way.\footnote{The FG expansion in this case only contains more power series terms with respect to the standard expansion. Those extra terms can be canceled one by one with local counterterms. It is possible to introduce an auxiliary scalar field to represent the extra contributions, ensuring that all solutions in the bulk are on-shell in the effective action sense, and to find the counterterms. The remaining finite part of the stress energy tensor is only related to the coefficient $g_{(d)}$ in the FG expansion.} The key elements allowing this treatment are the absence of logarithmic in $r$ terms in asymptotically AdS$_4$ spacetimes and that the extra power series divergence can be eliminated by introducing local counter-terms   \cite{Skenderis:2002wp,deHaro:2000vlm}. We will, therefore, track the finite coefficient in FG expansion as the boundary stress energy tensor\footnote{In the  Fefferman-Graham expansion, we have $ds^2 =\frac{d\rho}{4\rho^2}+\frac{1}{\rho} g_{ij}(x,\rho)dx^{i}dx^{j}$ and where $g(x,\rho)=g_{(0)}+\cdots +\rho^{\frac{d}{2}}g_{(d)}+\cdots $, and in 3d since there is no conformal anomaly, one can extract the coefficient $g_{(d)}$ to obtain $\langle T_{ij}\rangle =\frac{3}{16\pi G}g_{(d)ij}$.}
\begin{equation}
    T_{\mu\nu}=\frac{1}{2}\mathcal{E}\eta_{\mu\nu},\quad \mathcal{E}=\frac{1}{8\pi G}(4r_h^3 -4r_h Sch(u(v),v))\, .
\end{equation}
Substituting the saddle point value of $u(v)$, one obtains the boundary energy density  
\begin{equation}
    \mathcal{E}=\frac{1}{8\pi G}(4r_h^3 +8\pi^2 r_h T^2)\,,
\end{equation}
which gives the correct near-extremal RN black hole energy expression. The quantum average of $Sch(u(v),v)$ can be obtained by taking the derivative of the Schwarzian partition function, and we have \cite{Mertens:2017mtv}
\begin{equation}\label{Averaged Schwarzian derivative}
    \langle Sch(u(v),v)\rangle =-2\pi^2T^2 -\frac{3T}{2C}\, ,
\end{equation}
then the quantum-corrected boundary energy density is
\begin{equation}
\begin{aligned}
        \langle \mathcal{E}\rangle_{Sch} & = \frac{1}{8\pi G} \left(4r_h^3 - 4 r_h \langle Sch(u(v),v)\rangle \right) = \frac{1}{8\pi G} \left(4r_h^3 + 8\pi^2 r_h T^2 + \frac{6r_h T}{C}\right)
        \\ & = \frac{3r_h^2 C}{\pi} + 6\pi C T^2 + \frac{9}{2\pi}T,
\end{aligned}
\end{equation}
which is consistent with the result obtained directly from the partition function \cite{Iliesiu:2020qvm}. The entropy density of the fluid can be determined from the thermodynamic relation
\begin{equation}
    d\mathcal{E}=Tds +\mu d\rho\, .
\end{equation}
Since we are considering the canonical ensemble (fixed charge), one finally has

\begin{equation}
    s=s_0 +12\pi  \frac{T}{T_q}+\frac{9}{2\pi} \log \left(
\frac{T}{T_q} \right) \,.
\end{equation}
The results are up to an overall factor of $3/\pi$ compared with partition function results.\footnote{The overall factor can be understood from the difference between 2d and 4d holographic renormalization results. In general for odd boundary dimension $d$, we have $\langle T_{ij}\rangle=\frac{d}{16\pi G^{(d+1)}}g^{(d)}_{ij}$. Also, from dimensional reduction, we know that $G^{(2)}=G^{(4)}/V_{x,y}$, where $V_{x,y}$ is the volume of the compactified space, which has been set to $4\pi$.} In subsequent discussions, we will fix the charge of the black brane to $Q=\sqrt{3}r_h$ and focus on the canonical ensemble. When considering the linearized perturbation analysis, the charge density can fluctuate around the local equilibrium $r_h\to r_h+\delta r_he^{-i\omega v+ik_xx}$, but it remains globally conserved. Note also that since the chemical potential is defined as $\mu=\frac{Q}{r_{+}} $, in the very low temperature case, it satisfies $T\ll\mu\sim \sqrt{3}r_h$.

\subsection{Quantum-Corrected Equations of Motion: Fields in a Quantum Geometry}

After considering quantum corrections in thermodynamics, we now begin to consider quantum corrections in the fluid description (linearized perturbation). The upshot is that we will consider the equations of motion (EoMs) on the fluctuating bulk background and average the EoMs in the gravitational path integral as the quantum-corrected EoM, or the Schwinger-Dyson equation.

The standard approach for considering quantum corrections is ``back-reaction'' implemented by quantum averaging the matter stress tensor side of the Einstein equation and determining how quantum matter changes the metric:
\begin{equation}
    G_{\mu\nu}=\kappa \langle T_{\mu\nu}\rangle.
\end{equation}
A natural generalization of the back-reaction approach is to quantum average the full equation with the gravitational path integral
\begin{equation}
   \langle G_{\mu\nu}-\kappa T_{\mu\nu}\rangle =\int D[g_{MN}](G_{\mu\nu}-\kappa T_{\mu\nu})e^{-S[g_{MN}]}=0.
\end{equation}
We interpret the above equation as averaging an equation of motion (for matter on the background) with the metric.

Let us consider the example of a free massless scalar propagating in the AdS$_4$ black brane:
\begin{equation}
    \frac{1}{\sqrt{g}}\partial_{M}(\sqrt{g}\partial_{N}\phi\, g^{MN}) = 0\, .
\end{equation}

The most general way of including quantum corrections is to average the EoMs in the path integral. In some sense, the quantum-corrected equation of motion is the Schwinger-Dyson equation for one-point function in standard QFT, $\langle \frac{\delta S[\phi]}{\delta \phi}\rangle=0 $. In this consideration, we integrate out all metric degrees of freedom, i.e.,
\begin{equation}
    \left\langle\frac{1}{\sqrt{g}}\partial_{M}(\sqrt{g}\partial_{N}\phi g^{MN})\right\rangle = \frac{1}{Z}\int D[g_{MN}] \left(\frac{1}{\sqrt{g}}\partial_{M}(\sqrt{g}\partial_{N}\phi g^{MN})\right)\, e^{-S[g_{MN}]}\, .
\end{equation}
Since in the near-extremal limit, the action has been reduced to Eq.~\eqref{reducedaction}, the original degrees of freedom of $g_{MN}$ are also reduced to a 2d metric $g_{\mu\nu}^{(2d)}$ and a 2d linear dilaton $\Phi$. Now, $\Phi$ plays the role of a Lagrange multiplier, and $g_{\mu\nu}$ is fixed by $R=-2$. Consequently, we only need to integrate the boundary Schwarzian mode $u(v)$ as the leading contribution of quantum gravity effects. Moreover, we also need to perform a diffeomorphism transformation, which ensures that the excitation of the bulk geometry can be expressed in terms of the Schwarzian mode. Finally, the EoM of a massless scalar on this background is
\begin{equation}\label{Eq:modified metric}
    \frac{d}{dr} \left(6 \left((r-r_h)^2 + 2Sch(u(v),v)\right) \frac{d\phi(r)}{dr} \right) - 2i\omega \frac{d\phi(r)}{dr} + \frac{k_x^2}{r_h^2}\phi(r) = 0\, ,
\end{equation}
With this EoM, we can consider the quantum correction of a massless scalar as a prototype.

\section{Massless Scalar Model with Quantum Corrections}\label{sec:Massless Scalar Model}

In this section, we will discuss the massless scalar model as a prototype for incorporating quantum fluctuations of the metric in the treatment. In Sec.~\ref{sec:Review}, we have shown that taking the temperature to zero first leads to the presence of a non-local $\log(\omega)$ term. In this section, we introduce quantum corrections into the classical EoM, which will change the behavior of the solution, leading to a resolution of the $\log(\omega)$ problem.

\subsection{Quantum Corrections in Near-Extremal Case}

We will introduce quantum corrections to the massless scalar EoM and then calculate the first-order solution to the quantum-corrected EoM. We can firstly perform the diffeomorphism transformation in $AdS_2$ and express the finite energy excitation in terms of Schwarzian modes as
\begin{equation}
    ds^2=-6((r-r_h)^2+2Sch(u(v),v))dv^2 +2dvdr+r_h^2 (dx^2 + dy^2)\, .
\end{equation}
A crucial step is to align the NHR metric with the FAR metric and ensure that the quantum correction in NHR can be continuously transmitted to the asymptotic $AdS_4$ boundary. The FAR metric can be written as
\begin{equation}
    ds^2=-\frac{((r-r_h)^2+2Sch(u(v),v))(r^2 +2rr_h +3r_h^2)}{r^2}dv^2+2dvdr+r^2(dx^2 +dy^2)\,.
\end{equation}
Notice that this metric satisfies the asymptotic $AdS_4$ boundary condition and the continuous condition at the overlap region compared to the NHR metric. Let us highlight an important constraint that follows from the ability to match the AdS$_2$ boundary expansion with the asymptotic AdS$_4$ region. Consider the  near-horizon expansion of the metric factor $g_{tt}$
\begin{equation}
\begin{aligned}
        &\frac{((r-r_h)^2 +2Sch(u(v),v))(r^2 +2rr_h +3r_h^2)}{r^2}\\
        &\approx 12 Sch(u(v),v)-16Sch(u(v),v) \frac{r-r_{h}}{r_h}+2(3r_h^2 +11 Sch(u(v),v)) \left(\frac{r-r_h}{r_h}\right)^2 + \mathcal{O} \left( \left(\frac{r-r_h}{r_h}\right)^3 \right)\,.
\end{aligned}
\end{equation}
Note the appearance of a linear $\frac{r-r_h}{r_h}$ term. Since the saddle point value is $Sch(u(v),v)=-2\pi^2 T^2$, to ensure that the near-horizon asymptotic behavior matches the NHR metric, one should choose the overlap region as 
\begin{equation}
    T,\frac{1}{C}\,<(r-r_h)\ll\mu\,.
\end{equation}
Therefore, the $T^2 (r-r_h)$ and $T^2 (r-r_h)^2$ terms are higher-order terms for the near-horizon asymptotic expansion of the FAR metric.

\subsubsection{Quantum-Corrected Solution in NHR}

The EoM of a massless scalar in the fluctuating background can be written as
\begin{equation}
    \frac{d}{dr} \left(6 \left((r-r_h)^2 + 2Sch(u(v),v)\right) \frac{d\phi(r)}{dr}\right) - 2i\omega \frac{d\phi(r)}{dr} + \frac{k_x^2}{r_h^2}\phi(r) = 0\, ,
\end{equation}
and the quantum-corrected EoM is
\begin{equation}
\begin{aligned}
    \left\langle \frac{\delta I_{matter}}{\delta \phi} \right\rangle & = \frac{1}{Z_{RN}(\beta,r_h)}\int D[\Phi]\, D[g_{\mu\nu}^{(2d)}]\, e^{-I_{RN} [\Phi,g_{\mu\nu}^{(2d)}]-I_{matter}[\phi,g_{\mu\nu}^{(2d)}]} \frac{\delta I_{matter}}{\delta \phi}\\
    & = \frac{d}{dr} \left(6((r-r_h)^2 + \langle 2Sch(u(v),v)\rangle) \frac{d\phi(r)}{dr}\right) - 2i\omega \frac{d\phi(r)}{dr} + \frac{k_x^2}{r_h^2}\phi(r)\\
    & = \frac{d}{dr} \left(6 \left((r-r_h)^2 - 4\pi^2 T^2 - 3\frac{T}{C}\right) \frac{d\phi(r)}{dr}\right) - 2i\omega \frac{d\phi(r)}{dr} + \frac{k_x^2}{r_h^2}\phi(r)\\ 
    & = 0\,.
\end{aligned}
\end{equation}

In the high-temperature case ($T\gg \frac{1}{C}$), the leading term is $T^2$, which explicitly breaks the $AdS_2$ rescaling symmetry; the $\omega$-dependent term can be ignored. In the very low temperature case ($\frac{1}{C}\gg T$), the leading term is $\frac{T}{C}$ and since in the quantization of Schwarzian mode we have gauge fixed $SL(2,\mathbb{R})$, one can not perform time rescaling transformation to make $\omega$ important in NHR\footnote{The broken of $AdS_2$ symmetry has the following logic: At extremality, the pure $AdS_2$ has perfect conformal symmetry, we firstly choose a radial gauge and fix the boundary condition. The reparametrization symmetry is spontaneously broken to $SL(2,\mathbb{R})$ subset. Also, since we keep the finite excitation away from extremality, the symmetry is explicitly broken to an $SL(2,\mathbb{R})$ invariant effective  Schwarzian action. When considering the quantization of Schwarzian mode, one needs to do the path integral in $Diff(S^1)/SL(2,\mathbb{R})$ space, and $SL(2,\mathbb{R})$ symmetry as a gauge redundancy should be gauge fixed \cite{Stanford:2017thb}, therefore we can no longer perform time rescale transformation after quantum averaged Schwarzian mode.}. Therefore, with quantum correction, in both cases we can ignore the $\omega$-dependent term in the l.h.s, which is equivalent to considering 
\begin{equation}
    T<\omega <\frac{1}{C}<\mu\quad {\rm or} \quad \frac{1}{C}<\omega <T <\mu
\end{equation}
and have
\begin{equation}
        \frac{d}{dr} \left(6 \left((r-r_h)^2-4\pi^2 T^2 -3\frac{T}{C}\right) \frac{d\phi^{(1)}(r)}{dr}\right) = \frac{2i\omega\phi^{(0)}}{r_h} \, .
\end{equation}
The solution can be written as
\begin{equation}
    \phi^{(1)}_{in}(r)=A_{in}^{(1)}+\int_{r_+}^{r}\frac{dr'}{6((r'-r_h)^2 -4\pi^2 T^2 -3 \frac{T}{C})}(B_{in}^{(1)}+\int_{r_{+}}^{r'}\frac{2i\omega \phi^{(0)}}{r_h}dr'')\,,
\end{equation}
where we have set both lower limits of integration at the outer horizon $r_{+}=r_h+\sqrt{4\pi^2 T ^2 +3\frac{T}{C}}$. The solution is
\begin{equation}
\begin{aligned}
        \phi_{in}^{(1)} & = A_{in}^{(1)} + \frac{i\omega \phi^{(0)}}{6r_h} \Bigg[2\mathrm{arctanh} \left(\sqrt{\frac{C}{T(3+4 C \pi^2 T)}} (r-r_h) \right) + \log \left(-3T + C((r-r_h)^2 - 4\pi^2 T^2) \right)\\
        & \quad - i\pi - \log(4T(3 + 4C \pi^2 T))\Bigg].
\end{aligned}
\end{equation}
Expanding this solution near the overlap region, we have
\begin{equation}
\phi^{(1)}_{in} = A_{in}^{(1)} + \frac{i\omega \phi^{(0)}}{3r_h} \Bigg(\sqrt{\frac{C}{T(3+4\pi^2 C T)}}(r-r_h) -\log2+\cdots\Bigg)\, .
\end{equation}

\subsubsection{Quantum-Corrected Solution in FAR}

Now, we consider the FAR solution. The quantum-corrected EoM in FAR can be written as
\begin{equation}
    \frac{d}{dr} \left[ \left((r-r_h)^2 - 4\pi^2 T^2 - 3\frac{T}{C}\right) (r^2 +2rr_h +3r_h^2)\frac{d\phi^{(1)}}{dr}\right] - 2i\omega r \frac{d(r\phi^{(1)})}{dr}=2i\omega r \phi^{(0)}\, .
\end{equation}
We can ignore the $\omega$-dependent term in the l.h.s. Then the solution can be written as
\begin{equation}
\begin{aligned}
        \phi^{(1)}_{out}&=B_{out}^{(1)}-\int_{r}^{\infty}\frac{dr'}{r'^2f(r')_{FAR}}(A_{out}^{(1)}+\int_{r_{B}}^{r'}2i\omega \phi^{(0)}r'' dr'')\\
        &=B_{out}^{(1)}-\int_{r}^{\infty}\frac{dr'}{r'^2f(r')_{FAR}}(A_{out}^{(1)}+i\omega\phi^{(0)}(r^2-r_B^2))\,.
\end{aligned}
\end{equation}
We have set the inner integration limit to the overlap region location $r_B$, and by using the normalizable condition $\phi^{(n)}=0$ for $n>1$ when $r\to \infty$, we can set $B_{out}^{(1)}=0$. For simplicity, we denote $ \delta r_{+}=r_+-r_h =\sqrt{4\pi^2 T^2 +3T/C}$. Then, near the overlap region, the asymptotic behavior of the outer solution can be written as
\begin{equation}
\begin{aligned}
       \phi^{(1)}_{out}= Constant -\frac{A_{out}^{(1)}-i(r_B^2 -r_h^2)\phi^{(0)}\omega}{6r_h^2 \delta r_{+}^2}(r-r_h) + \mathcal{O} \left( \left(\frac{r-r_h}{r_h}\right)^2 \right)\,.
\end{aligned}
\end{equation}
Comparing the coefficient of the $(r-r_h)$ term with the previous expansion of the NHR solution, one can fix the integration constant $A_{out}^{(0)}$. However, naively match the coefficient yields $A_{out}^{(1)}=i\omega \phi^{(0)}(r_B^2 -r_h^2 +2 \delta r_{+})$, which is equivalent to extend the inner integration limit from $r_B$ to $\sqrt{r_h^2 +2 \delta r_{+}}$. It is a proper choice, but the physical picture is unclear as to why the integration should end at this location. A more natural choice is 
\begin{equation}
    A_{out}^{(1)}=i\omega \phi^{(0)}(r_B^2 -(r_h +\delta r_{+})^2)\, .
\end{equation}
This matching choice has the clear physical interpretation of extending the inner integration limit from $r_B$ to $r_h+\delta r_{+}=r_{+}$, which is the out-most horizon location. The expansion of the outer solution becomes
\begin{equation}
    \phi^{(1)}_{out}=Constant +\frac{i\omega \phi^{(0)}(r-r_h)}{3\delta r_{+}r_h}+\frac{i\omega \phi^{(0)}(r-r_h)}{6r_h}+\cdots
\end{equation}
As we can see, the leading term matches with the inner solution, and the following term is subleading in the expansion regime of the overlap region($T,\frac{1}{C}<(r-r_h)\ll \mu$). Therefore, this match choice for the FAR solution is valid when near the overlap region. The full solution in the FAR is
\begin{equation}
\begin{aligned}
       \phi^{(1)}_{out} & = \frac{\omega\phi^{(0)}}{2r_h (6r_h^2 -4 r_h \delta r_{+}+\delta r_{+}^2)} \Bigg[2\pi r_h^2 +i\sqrt{2} (2r_h^2 -2 r_h\delta r_{+}+\delta r_{+}^2) \arctan \left(\frac{\sqrt{2}r_h}{r+r_h}\right)\\
       & \quad + 2ir_h^2 \left(2\mathrm{arctanh} \left(\frac{r_h-r}{\delta r_{+}}\right) + \log \Big(\frac{r^2 +2 r r_h +3 r_h^2}{(r-r_h)^2 -\delta r_{+}^2}\Big) \right)\Bigg]\\
       & = \frac{\omega\phi^{(0)}}{2r_h (6Cr_h ^2 +3 T +4\pi^2  CT^2 - 4 r_h \sqrt{CT (3 +4\pi^2  C T)} )} \Bigg[2C\pi r_h^2 +i \sqrt{2} \Big(3T \\
       & \quad -2r_h \sqrt{C T (3+4 C\pi^2 T)} + 2C (r_h^2 +2\pi^2 T^2) \Big) \arctan \Big(\frac{\sqrt{2}r_h}{r+r_h}\Big)\\
       & \quad + 2i C r_h^2 \left(2 \mathrm{arctanh}\Big(\sqrt{\frac{C}{T(3+4 C \pi^2 T)}}(r_h-r)\Big)+\log\Big(\frac{r^2 +2rr_h +3r_h^2}{(r-r_h)^2 -\frac{3T}{C}-4\pi^2 T^2}\Big) \right)\Bigg]\,.
\end{aligned}
\end{equation}
By matching the solution near the overlap region, the integration constant $A_{in}^{(1)}$ in the inner solution can also be fixed to
\begin{equation}
\begin{aligned}
    A_{in}^{(1)} & = -\frac{i\omega \phi^{(0)}}{(6\sqrt{2}r_h (\delta r_{+}^2 -4\delta r_{+}r_h +6r_h^2))}\Bigg(-3\pi(\delta r_{+}-2\delta r_{+}r_h +2r_h^2)+6(\delta r_{+}^2 -2 \delta r_{+}r_h \\
    & \quad + 2 r_h^2)\arctan(\sqrt{2})
    +\sqrt{2}\delta r_{+}(8r_h \log2-2\delta r_{+}\log2) -6\sqrt{2}r_h^2 \log \left(\frac{24 r_h^2}{\delta r_+^2} \right)\Bigg)\,.
\end{aligned}
\end{equation}
Both the integration constant in NHR and FAR solutions have been completely fixed. As we can see, there is no $\log(\omega)$ term, which makes the fluid derivative expansion valid.

From this prototype, we can observe that the matching process is equivalent to extending the outer solution's integration limit from $r_{B}$ to $r_{+}$. Therefore, in the later discussion of metric and gauge field perturbation, we will apply this observation and directly set the inner integration limit to the outer horizon $r_{+}$.

\subsection{The Non-Analytic Term $\log(\omega)$ and Proper Regime Choice}\label{sec: non-analytic term}

In this subsection, we will discuss the reason why non-analytic terms appear and the proper choice of expansion regime that eliminates them. The main observation is that in the quantum-corrected EoM of NHR
\begin{equation}
    \frac{d}{dr} \left[6 \left((r-r_h)^2-4\pi^2 T^2 -3\frac{T}{C}\right) \frac{d\phi^{(1)}(r)}{dr} \right] - 2i\omega \frac{d\phi^{(1)}(r)}{dr}=\frac{2i\omega\phi^{(0)}}{r_h} \, ,
\end{equation}
the $\omega$-dependent term in the l.h.s is always subleading. If the $\omega$-dependent term is leading and can not be ignored, the $\log(\omega)$ will appear in the solution. 

The equation above can be rewritten as
\begin{equation}
    \frac{d^2\phi^{(1)}}{dr^{*2}}-2i\omega \frac{d\phi^{(1)}}{dr^*}=\frac{2i\omega \phi^{(0)}}{r_h}f(r)_{NHR}\, ,
\end{equation}
with $r^*=\int \frac{dr}{f(r)_{NHR}}=-\frac{\mathrm{arctanh}(\frac{r-r_h}{\delta r_+})}{6\delta r_{+}}-\frac{i\pi}{12 \delta r_+}$, and the solution can be written in the form of 
\begin{equation}
    \phi^{(1)}_{in}=A_{in}^{(1)}+B_{in}^{(1)}e^{2i\omega r^*}+\frac{2i\omega \phi^{(0)}}{r_h}e^{2i\omega r^{*}}\int_{-\infty}^{r^{*}}d{r^*}'e^{-2i\omega {r^*}'}\int_{r_+}^{r'}dr''\,.
\end{equation}
With the ingoing boundary condition at horizon we can set $B_{in}^{(1)}=0$, directly integrate gives
\begin{equation}
    \phi^{(1)}_{in}=A_{in}^{(1)}+\frac{2i\omega \phi^{(0)}}{r_h}\cdot \frac{i}{2\omega} \left(-\delta r_+ +r_h +2 \delta r_+{_2 F_{1}(1,-\frac{i\omega}{6 \delta r_{+}},1-\frac{i\omega}{6 \delta r_+},e^{12 \delta r_{+}r^{*}})} \right)\, .
\end{equation}
One can analyze the asymptotic expansion of the hypergeometric function to obtain the behavior of the solution. However, we can perform this in a more transparent way, following the method of \cite{Moitra:2020dal}, in which we denote the integration as
\begin{equation}
    I^{(1)}_{in}(r^{*};T,\omega)=\int_{-\infty}^{r^*}{dr^*}' e^{-2i\omega {r^*}'}\Bigg(r_h-\delta r_{+} \tanh \Big(\frac{1}{2}(i\pi +12 \delta r_{+}{r^{*}}')\Big)\Bigg)\,.
\end{equation}
We can take the derivative of this integration with respect to $\omega$ and separate to have
\begin{equation}
\begin{aligned}
        \frac{\partial I_{in}^{(1)}(r^{*};T,\omega)}{\partial \omega} & = \int_{-\infty}^{0}d{r^{*}}' e^{-2i\omega {r^*}'}(-2i{r^{*}}')\Bigg(r_h-\delta r_{+} \tanh \Big(\frac{1}{2}(i\pi +12 \delta r_{+}{r^{*}}')\Big)\Bigg)\\
        & \quad + \int_{0}^{r^{*}}{dr^{*}}' e^{-2i\omega {r^*}'}(-2i{r^{*}}')\Bigg(r_h-\delta r_{+} \tanh \Big(\frac{1}{2}(i\pi +12 \delta r_{+}{r^{*}}')\Big)\Bigg)\, .
\end{aligned}
\end{equation}
For the first integration, it can be done directly and yields a polygamma function, and for the second, we will expand in small $\omega r^{*}$, then perform the integration, and then the results give
\begin{equation}
    \begin{aligned}
        \frac{\partial I_{in}^{(1)}(r^{*};T,\omega)}{\partial \omega} & = 2i \left(\frac{\delta r_{+} - r_h}{4\omega^2} + \frac{\psi^{(1)}(-\frac{i\omega}{6\delta r_{+}})}{72 \delta r_{+}}\right) + \frac{1}{9}r^{*} \Big(3i + r^* \big(12 i \delta r_{+}^2 r^{*} 
        \\& \quad + \omega (3-2i\omega r^{*})-3r_h(3i +4 \omega r^*) \big)\Big) + \cdots\,.
    \end{aligned}
\end{equation}
Integrating $\omega$ can yield
\begin{equation}
\begin{aligned}
    I_{in}^{(1)}(r^* ; T, \omega)=I_{c}(r^*;T) - \frac{1}{6} \psi \left(-\frac{i\omega}{\delta r_{+}}\right) - \frac{i(\delta r_{+}-r_h)}{2\omega} + \frac{1}{3}i(1+12 \delta r_+^2 {r^{*}}^2 -6r_h r^{*})\omega + \mathcal{O}(\omega^2, r^{*})\,
\end{aligned}
\end{equation}
where $I_{c}(r^*;T)$ is an integration constant which does not contain $\omega$. The polygamma function plays a central role in clarifying the regime in which $\log(\omega)$ appears, a phenomenon widely observed in the literature \cite{Moitra:2020dal, Gouteraux:2025kta}. Now we consider the behavior of the polygamma function in different regimes,
\begin{equation}
    \psi \left(-\frac{i\omega}{\delta r_{+}}\right) = \psi \left(-\frac{i\omega}{\sqrt{4\pi^2 T^2 +3T/C}}\right)\, .
\end{equation}
\begin{itemize}
\item[a)]  The classical low-temperature limit: Scenario 1 ($T_q < T\ll \omega \ll \mu$)

In this regime, the Schwarzian scale $T_q$ is the smallest energy scale, and quantum fluctuations provide a tiny correction around the classical value. We have
\begin{equation}
    \begin{aligned}
        \psi \left(-\frac{i\omega}{\sqrt{4\pi^2 T^2 +3TT_q}}\right) & \approx \psi \left(-\frac{i\omega}{2\pi T}\right) + \frac{3i\omega T_q}{16\pi^3 T^2 }\, \psi^{(1)} \left(-\frac{i\omega}{2\pi T}\right) + \cdots\\
        &\approx-\frac{3T_q}{8\pi^2  T}+\frac{3iT_q+8\pi \omega}{8\pi \omega} \log \left(-\frac{i\omega}{2\pi T}\right) + \cdots\,.
    \end{aligned}
\end{equation}
We can see that in this regime, the $\log(\omega)$ term will appear.

\item[b)] The classical low-temperature limit: Scenario 2 ($T_q\ll \omega \ll T \ll \mu$)\\
It is easy to show that
\begin{equation}
    \begin{aligned}
        \psi \left(-\frac{i\omega}{\sqrt{4\pi^2 T^2 +3TT_q}}\right) & \approx \psi \left(-\frac{i\omega}{2\pi T}\right) + \frac{3i\omega T_q}{16\pi^3 T^2}\, \psi^{(1)} \left(-\frac{i\omega}{2\pi T}\right) + \cdots\\
        & \approx \left(-\frac{3iT_q}{4\pi } - 2i\pi T\right) \frac{1}{\omega}-\gamma -\frac{i(-3T_q+8\pi^2  T)\omega}{96 \pi T}\\
        & \quad + \frac{-3T_q+4\pi^2 T}{16 \pi^4  T^3}\zeta(3) \omega^2 +\cdots \,,
    \end{aligned}
\end{equation}
where $\zeta(3)$ is the Riemann zeta function. As we can see, in this regime, the $\log(\omega)$ term disappeared.  For those 2 cases, one could observe a leading contribution as $\psi(-\frac{i\omega}{2\pi T})$ which is similarly observed in the retarded Green's function of \cite{Gouteraux:2025kta}, and the regime choice of whether $T<\omega$ or $\omega <T$ is crucial to determine the existence of non-analytic $\log(\omega)$ term and shows an instability of the fluid derivative expansion. Besides, in the latter two cases, we will show that by including a quantum correction, the leading part of the argument in the polygamma function will change, and the $T_q$ can play a role in stabilizing the relaxation process of the fluid perturbation.

\item[c)]The quantum low-temperature limit: Scenario 1 ($T\ll T_q\ll \omega \ll \mu$)

In this regime, the Schwarzian fluctuations become dominant with $TT_q>T^2$. Now the polygamma function behaves like

\begin{equation}
    \begin{aligned}
         \psi \left(-\frac{i\omega}{\sqrt{4\pi^2 T^2 +3TT_q}}\right) & \approx \psi \left(-\frac{i\omega}{\sqrt{3TT_q}}\right) + \cdots
         \\ & \approx \log \left(-\frac{i\omega}{\sqrt{3TT_q}}\right) - \frac{i\sqrt{3TT_q}}{2\omega} + \frac{3T_q^2 - 8\pi^2\omega^2}{12 \omega^2 T_q}T + \cdots\,.
    \end{aligned}
\end{equation}
As we can see, in this regime even with the quantum correction, the $\log(\omega)$ term still appears\footnote{It should be noticed that, when $\omega >\frac{1}{C}$, it easy to show that $\omega >\sqrt{T_q^2}>\sqrt{TT_q}$. However, for $\omega <T_q$, we should still put $\sqrt{TT_q}>\omega >\sqrt{T^2}$ which is the $T_q$ very strong limit, to guarantee the $\log(\omega)$ will disappear.}.

\item[d)]  The quantum low-temperature limit: Scenario 2 ($T< \omega < T_q \ll \mu$)\\
In this regime, we have

\begin{equation}
    \begin{aligned}
        \psi \left(-\frac{i\omega}{\sqrt{4\pi^2 T^2 + 3T T_q}}\right) & \approx \psi \left(-\frac{i\omega}{\sqrt{3TT_q}}\right) + \cdots
         \\& = -\frac{i}{\omega}\sqrt{3TT_q}-\gamma -\frac{i\pi^2 (12 T^2 +\omega^2)}{6\omega}\sqrt{\frac{1}{3TT_q}}+\frac{\omega^2 }{3TT_q}\zeta(3)+\cdots \,.
    \end{aligned}
\end{equation}

In this regime, with quantum correction and proper regime choice, the $\log\omega$ term disappears and the non-analytic problem of the fluid derivative expansion can be resolved. 
\end{itemize}
Let us summarize the situation regarding the various orders of limits in the  low-temperature fluid/gravity correspondence: 

    \begin{equation}
        \begin{aligned}
\text{Semi-Classical}:&\left\{\begin{matrix}
T_q<T<\omega <\mu\, , & \log\omega\, \text{ exists;}\\
T_q<\omega <T <\mu\, , &\log \omega\, \text{ vanishes;}
\end{matrix}\right.\\
\text{Near-extremal}:&\left\{\begin{matrix}
T<T_q<\omega<\mu\, , & \log\omega\, \text{ exists;}\\
T<\omega<T_q<\mu\, , &\log \omega\, \text{ vanishes.}
\end{matrix}\right.
\end{aligned}
    \end{equation}
    
In principle, fluid dynamics provides an effective description when a field theory reaches local equilibrium, and the original degrees of freedom can be reduced to effective fluid quantities, such as temperature, velocity, and chemical potential. In the conventional fluid gravity correspondence, we choose $\omega,k_x \ll T,\mu$, where the perturbation is small and cannot break the equilibrium. Intuitively, allowing $T\ll\omega$ implies that the perturbation scale $(\omega)$ is much bigger than the equilibrium scale $(T)$, which means a single perturbation will break the equilibrium and make the system unstable, leading potentially to a non-equilibrium regime.

This is our physical explanation for the appearance of  $\log({\omega})$ terms.  When quantum fluctuations are the leading effect ($T_q >T$), if perturbation energy is in the window $T<\omega <T_q$, a perturbation might break the thermal equilibrium at first since $T<\omega$, but with $\omega<T_q$, the energy is not sufficient to generate higher energy modes than the Schwarzian mode. Therefore, the leading quantum fluctuation will help the system quickly relax to local equilibrium again, and the system is robust against low-energy perturbations ($\omega < T_q$). In other words, the Schwarzian quantum fluctuation plays a role in stabilizing the system's relaxation process at low temperatures, preventing the system from becoming out of equilibrium. 
\begin{figure}[h]
    \centering
    \includegraphics[width=0.8\textwidth]{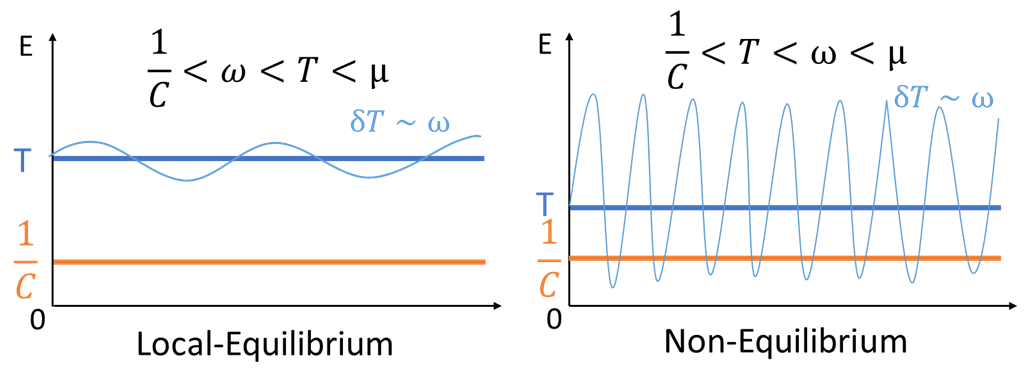}
    \caption{Regime Choice and Local-Equilibrium}
\end{figure}

Besides the physical interpretation of why $\log(\omega)$ appears, the technical explaination is: when we choose different regime, the leading terms in the equation of motion will change, and only when we can omit the $\omega\frac{d\phi}{dr}$ term as a subleading term, the $\log\omega$ problem could be solved.

Let us rephrase the result in the language of symmetry. The original reason for keeping the $\omega\frac{d\phi}{dr}$ term in the EoM is the scaling symmetry of near-horizon $AdS_{2}$, which implies that the $\omega$-dependent term in EoM is always important. By choosing a different regime, the rescaling symmetry can either be explicitly broken or gauge-fixed. Then, for the derivative expansion, the $\omega$-dependent term is always suppressed. These can be illustrated in both EoM level ($\omega$-dependent term in the l.h.s) and solution level (Asymptotic expansion of polygamma function $\psi(-\frac{i\omega}{\delta r_{+}})$).

 Previously it has been shown that the $\log(\omega)$ only appears in the first order inner solution's integration coefficients, which can be shown that will growth to $A_{out}^{(n)}\sim\epsilon^{n}(\log(\frac{\omega}{r_h}))^{n-1}$\cite{Moitra:2020dal}. Although here we only illustrate how the proper regime choice and quantum correction can resolve the $\log(\omega)$ problem in the first order calculation, since this guarantees the $\omega$-dependent term can always be ignored in the l.h.s of EoM, the conclusion here can be simply generalized to the higher order calculation for 
\begin{equation}
    \frac{d}{dr} \left(r^4 f(r)Q_{cor}(r)\frac{d\phi^{(n)}}{dr}\right) = s^{(n)}\,.
\end{equation}
Where $s^{(n)}$ is a source term of $\mathcal{O}(\epsilon^n)$ determined by the solution up to the previous orders $\phi^{(n-1)},\cdots, \phi^{(1)},\phi^{(0)}$. Since there is no $\omega$-dependent term in the l.h.s, the order of derivative expansion on each side of EoM can be well matched, and the $\log(\omega)$ terms will not appear in higher order either.

\section{Quantum Corrections to Near-Extremal Fluid/Gravity Correspondence}\label{sec:Near-Ext Fluid Gravity Correspondence}

In this section, we will extend the method discussed in the massless scalar model to the metric and Maxwell field perturbations.

\subsection{Zeroth-Order Stress Tensor and Charge Current for Boundary Fluid}

We will first calculate the quantum-corrected zeroth-order perturbations of the metric and gauge fields. Subsequently, via holographic renormalization, we will obtain the zeroth-order stress tensor and charge current for the boundary fluid.

As a first step, we turn on perturbations by promoting the fluid dynamics quantities to depend on the boundary coordinates,
\begin{equation}\label{adding coordinates dependence in fluid quantities}
    \begin{aligned}
           r_{h}(x^{\sigma}) & = r_{h} + \delta r_{h}\, e^{-i\omega v + ik_{x} x}\, ,\\
           T(x^{\sigma}) & = T + \delta T\, e^{-i\omega v + ik_{x} x}\, ,\\
           u_{\mu}(x^{\sigma}) & = (-1,\, \delta \beta_{x}\, e^{-i\omega v + i k_{x}x},\, \delta \beta_{y}\, e^{-i\omega v + ik_{x}x})\, ,
    \end{aligned}
\end{equation}
where $\delta r_{h}$, $\delta T$, and $\delta \beta _{i}$ are constants independent of $x^{\mu}$. Moreover, $r_{h}$ represents the chemical potential $\mu$ since $\mu \approx \sqrt{3} r_{h}$. We have chosen the local rest frame for the fluid, which means that the fluid has zero background velocity in this frame, and the perturbations are just a small local deformation. We work in the regime where all the deformations $\delta T$, $\delta \beta_{i}$, and $\delta r_{h}$, are of the  same order
\begin{equation}
    \delta T \sim \delta \beta_{i} \sim \delta r_{h}\, .
\end{equation}
Since we aim to examine the linearized perturbative behavior, in subsequent calculations, we will retain only the first order in the variation parameters $\delta T$, $\delta \beta_{i}$, and $\delta r_{h}$. Recall that the expansion with respect to the variation parameters and the derivative expansion with respect to $T$, $\omega$, and $k_{x}$ should be taken separately. The variation is a characteristic of the perturbation scale we add to this system, but the $T$-, $\omega$-, and $k_{x}$-scales are intrinsic and only depend on the system itself.

We can now proceed to implement the substitution Eq.~\eqref{adding coordinates dependence in fluid quantities} into the quantum-corrected background metric Eq.(\ref{Eqn:BackgroundMetric}) in which we use the quantum averaged Schwarzian derivative Eq.(\ref{Averaged Schwarzian derivative}) to obtain the linear temperature term $\frac{T}{C}$ and the $T^2$ term comes from classical value of Schwarzian derivative
\begin{equation}
    \begin{aligned}
        ds^2 &=\bar{g}_{MN}dx^{M}dx^{N}=-\frac{((r-r_h)^2 +2\langle Sch(u(v),v)\rangle)(r^2 +2rr_h +3r_h^2)}{r^2}dv^2+2dvdr+r^2(dx^2 +dy^2)\\
        &=-\frac{((r-r_h)^2 -4\pi^2 T^2 -3T/C)(r^2 +2rr_h +3r_h^2)}{r^2}dv^2+2dvdr+r^2(dx^2 +dy^2)\\
        A_{M}&=\bar{A}_{M}=-\frac{\sqrt{3}r_h^2}{r}dv.
    \end{aligned}
\end{equation}
Decomposing the zeroth-order metric into the background part and the perturbation part leads to 
\begin{equation}\label{eq:perturbed metric}
\begin{aligned}
  ds^2 & = \bar{g}_{MN}dx^{M}dx^{N}+e^{-i\omega v+ik_{x}x }h_{MN}(r)dx^{M}dx^{N}\, ,\\
  A_{M} & = \bar{A}_{M}+e^{-i \omega v+i k_{x} x}a_{M}(r)\, .
\end{aligned}
\end{equation}
From the background and zeroth order perturbation we obtained (See App.~\ref{App:Pert-Details}), the zeroth order stress tensor and charge current 
\begin{equation}
\begin{aligned}
        T_{\mu\nu}^{(0)}&=\frac{1}{2}\mathcal{E}(\eta_{\mu\nu}+3u_{\mu}u_{\nu})\\
    J_{\mu}^{(0)}&=\rho u_{\mu}\,,
\end{aligned}
\end{equation}
where the energy density is
\begin{equation}
    \mathcal{E}=\frac{1}{8\pi G}\Bigg (4r_h^3 +8\pi^2 r_h T^2 + \frac{6r_h T}{C} +\Big( 4(3r_h^2 \delta r_h +2 \pi^2 T^2 \delta r_h +4\pi^2 r_h T \delta T)+\frac{6(T\delta r_h +r_h \delta T)}{C}\Big)e^{-i\omega v +ik_x x}\Bigg)\,,
\end{equation}
and the charge density is
\begin{equation}
    \rho =\frac{\sqrt{3}}{8\pi G}(2r_h^2 +4r_h \delta r_h e^{-i\omega v +ik_x x})\, .
\end{equation}
We can see from the energy density that the $T^2$ term represents the semi-classical near-extremal temperature correction and the $\frac{T}{C}$ term comes from quantum correction, where the crucial linear dependence of $T$ in energy density can address the problem of the semiclassical breakdown of thermodynamics and provides $\log{T}$ in the entropy density.. 

Such constructed zeroth-order stress-energy tensor and charge current satisfy the conservation equations
\begin{equation}
    \partial_{\mu} T^{\mu\nu} = 0\, ,\quad \partial_{\mu} J^{\mu} = 0\, ,
\end{equation}
which describe a dual ideal fluid dynamics on the 3-dimensional boundary. In the following subsections, we will construct the dissipative sector of the boundary fluid with the bulk first-order perturbation results.

\subsection{Linearized Perturbation for Metric and Gauge Field}

The EoM for the metric and the gauge field perturbations can be generally written as
\begin{equation}
    \mathbb{D}(g^{(n)}_{MN},\, A^{(n)}_{M}) = s^{(n)} (g^{(n-1)}_{MN},\, \cdots,\, A_{M}^{ (n-1)},\, \cdots)\, ,
\end{equation}
where the differential operator $\mathbb{D}$ acts on the pertubations at the $n-$th order. The r.h.s indicates that the source of the $n$-th order perturbative solution is determined by the $(n-1)$-th solution. We will first derive the classical EoM for the perturbation and then take the quantum average of the EoM. The classical EoM for the linearized perturbation remains the same as \cite{Moitra:2020dal}, which can be obtained by introducing some linearized perturbations on the background metric and gauge field, where the background metric and the background gauge field take their classical values:
\begin{equation}
\begin{aligned}
    \bar{g}_{MN}\, dx^{M} dx^{N} & = -r^2 f(r)\,dv^2 + 2 dv\, dr + r^2 (dx^2 + dy^2)\, ,\\
    \bar{A}_{M}\, dx^{M} & = g(r)\, dv\, .
\end{aligned}
\end{equation}
By working with the gauge fixing condition
\begin{equation}
    h_{rr} = 0 = a_{r}\, ,
\end{equation}
we plug the Ansatz for the linearized perturbations, Eq.~\eqref{eq:perturbed metric}, back to the original Einstein-Maxwell equations and only keep the perturbations $h_{MN}(r)$ and $a_{M}(r)$ to the linear order, then we obtain the linearized perturbation EoM for the metric and the gauge field. It is convenient to raise an index of the gravitational perturbation, $h_{MN}$, using the background metric ${h^{M}}_{N} = \bar{g}^{MS} h_{SN}$, and denote ${h^{x}}_{x} = \sigma(r) + \alpha(r) $, ${h^{y}}_{y} = \sigma(r) - \alpha(r)$. With the classical EoMs, we can first take the quantum average in the NHR and then extend the solution to the FAR, with each metric factor in the classical EoM receiving a $Q_{cor}$ correction factor. 
We will first divide the perturbations into the $SO(2)$ tensor, vector, and scalar sectors:

\begin{itemize}
    \item Tensor sector: $g_{xy}^{(n)}$, $(g_{xx}^{(n)}-g_{yy}^{(n)})/2$
    \item Vector sector: $g_{vi}^{(n)}$, $A_{i}^{(n)}$
    \item Scalar sector: $g_{vv}^{(n)}$, $g_{vr}^{(n)}$, 
    $(g_{xx}^{(n)}+g_{yy}^{(n)})$,
    $A_{v}^{(n)}$
\end{itemize}
The perturbations in the same sector will obey EoMs with a similar form. The calculation details of each metric and gauge field component at first order in perturbation can be found in App.~\ref{App:Pert-Details}.

\subsection{First-Order Stress Tensor and Charge Current for Boundary Fluid}

In this subsection, we use the first-order perturbation results to calculate the first-order stress tensor and charge current for the boundary fluid. The details of the first-order perturbation theory can be found in App.~\ref{App:Pert-Details}. Using the same method as for the zeroth-order case, we obtain the first-order stress tensor in the form:
\begin{equation}
T_{\mu\nu}^{(1)} =-\eta e^{-i\omega v + ik_{x}x} \begin{pmatrix}
  T_{vv}^{(1)} & T_{vx}^{(1)} & T_{vy}^{(1)} \\
  T_{xv}^{(1)} & ik_{x}\delta\beta_{x} & ik_{x}\delta \beta_{y} \\
  T_{yv}^{(1)} & ik_{x}\delta\beta_{y} & -ik_{x}\delta\beta_{x} 
\end{pmatrix}\, ,
\end{equation}
where the shear viscosity $\eta$ in the $T<T_q$ regime is
\begin{equation}
\begin{aligned}
    \eta=\frac{r_h^2}{16\pi G}+\frac{\sqrt{3}r_h \sqrt{T}}{8\pi G \sqrt{C}}-\frac{15T}{8\pi G C}+\frac{(36+C^2 \pi^2 r_h^2)T^{3/2}}{4\sqrt{3}\pi G C^{3/2}r_h} -\frac{(20\pi ^2 +\frac{99}{C^2 r_h^2})T^2}{8\pi G}+\mathcal{O}(T^{5/2})\,.
\end{aligned}
\end{equation}
For comparison, if we expand the first-order perturbation in the limit of $T>T_q$, we obtain
\begin{equation}
    \begin{aligned}
        \eta&=\frac{1}{16\pi G}\Big(\frac{r_h(r_h^3 +8\pi r_h^2 T -20 \pi^2 r_h T^2 -16 \pi^3 T^3)}{(r_h +2\pi T)^2}-\frac{r_h(-r_h^3 +14\pi r_h^2 T +12\pi^2 r_hT^2+8\pi^3 T^3)}{2\pi (r_h +2\pi T)^3 C}+\cdots\Big)\\
        &\approx \frac{r_h^2}{16\pi G} +\frac{r_hT}{4G}+\frac{3r_h}{32\pi^2 C G}-\frac{15 T}{8\pi C G }+\mathcal{O}(T^2,\frac{1}{C^2})\,.
    \end{aligned}
\end{equation}
As we can see, in the $T>T_q$ limit, the leading order of shear viscosity is proportional to the area of outermost horizon ($A_{H}\propto r_{+}^2=r_h^2+4\pi r_h T +4\pi^2 T^2 $), which is consistent with the semi-classical results \cite{Policastro:2001yc}.

\begin{figure}[h]
    \centering
    \includegraphics[width=0.8\textwidth]{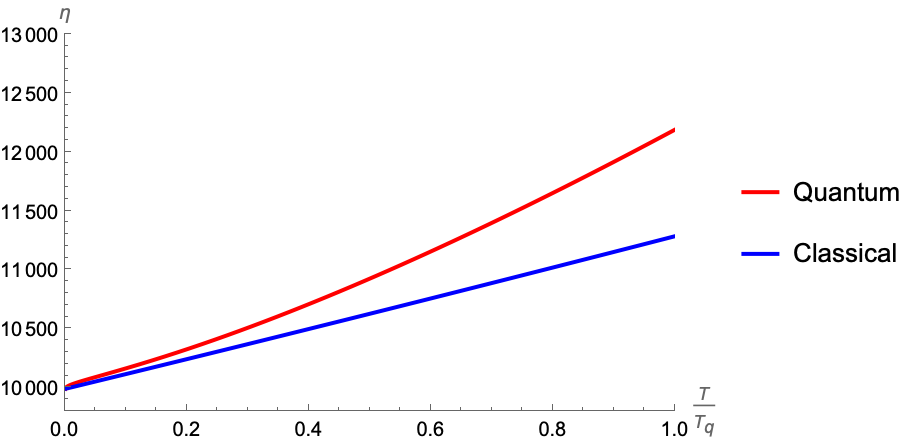}
    \caption{Temperature dependence of $\eta$ at $C=1,\,r_h=100$. The red line represents the quantum-corrected $\eta$ in the $T<T_q$ limit, and the blue line represents the semi-classical area law of $\eta$.}
    \label{class-quant eta}
\end{figure}

As shown in Fig.~\ref{class-quant eta}, in the $T<T_q$ regime, the quantum-corrected $\eta$ becomes larger than the classical area law $(\eta \propto A_{H})$. This behavior is consistent with a recent discussion advocating for an enhancement of the quantum absorption cross section $\sigma_{abs}$ \cite{Emparan:2025sao}. Since our expansion is in the regime of $T\ll T_q$, the expansion results will be reliable when temperature is very low but become less reliable towards $T\sim T_q$. While averaging in the metric, we only kept order $\mathcal{O}(T^2)$ terms as shown in \eqref{Eqn:BackgroundMetric} and, subsequently, we limit our plots to include corrections to that same order. The components  $T_{vv}^{(1)}, T_{vi}^{(1)}$ can be set to zero by using the Landau frame condition (see App.~\ref{sec: landau frame condition}). The one-loop corrected entropy density equals 
\begin{equation}
\begin{aligned}
        s&=\frac{r_h^2}{4G}+\frac{2\pi r_h T}{G}+\frac{3r_h}{4\pi CG}\log(CT)\\
        &=\frac{3r_hC}{2}+12\pi CT+\frac{9}{2\pi}\log(CT)\,,
\end{aligned}
\end{equation}
where we have used $G=\frac{r_h L_2^2}{C}=\frac{r_h L^2}{6C}=\frac{r_h}{6C}$. Positivity of the entropy density clarifies the regime of validity of the above approximation by implying  a lower bound for temperature
\begin{equation}\label{Eq:T0}
   s|_{T\to T_{0}}\approx 0 \quad\Longrightarrow\quad T_{0}\approx \frac{1}{C}e^{-\frac{1}{3}\pi r_h C}\,,
\end{equation}
where since $\frac{1}{C}\ll r_h$, we have $Cr_h \gg 0$. 

We illustrate the temperature dependence of $\eta/s$ by plotting it in the regime $T<\frac{1}{C}<r_h$. 
\begin{figure}[h]
    \centering
    \includegraphics[width=0.7\textwidth]{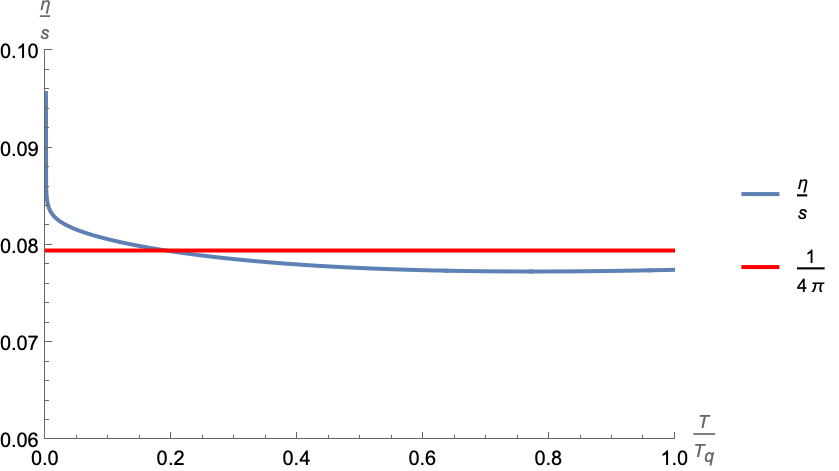}
    \caption{Temperature dependence of $\eta/s$ with $C=1\,,r_h=100$.}
    \label{eta/s 1/4pi}
\end{figure}
For extremely low temperatures, approaching $T_0$ above, the value of $\frac{\eta}{s}$ diverges, indicating the edge of the regime of validity. Interestingly, we find a violation of the $1/4\pi$ bound in the regime where the temperature is non-zero but vanishingly small.  Namely, the value of $\eta/s$  becomes lower than the $1/4\pi$ bound because of the breaking of the NHR conformal symmetry. We emphasize that this result is possible only in the quantum regime, $T<\frac{1}{C}$.

For the low-temperature regime, we can have
\begin{equation}\label{eta/s expansion}
    \frac{\eta}{s}\approx \frac{r_h/T_q}{4\pi  r_h/T_q +12 \log(\frac{T}{T_q})}+\frac{\sqrt{3}\sqrt{T/T_q}}{2\pi r_h/T_q +3\log(\frac{T}{T_q})}+\cdots\,.
\end{equation}

To explore other corners of the shear viscosity to entropy density ratio, it is crucial to understand the order of limits. For example, for a fixed, non-zero temperature, $T$, we can then take the large chemical potential limit, large $\mu$; the answer in this order of limits leads to $1/4\pi$ which recovers \cite{Davison:2013bxa} with the crucial caveat that the limit must be taken with the temperature, while small, can not be taken all the way to zero. Indeed, taking $T\to 0$ first pushes us to the regime of validity of the computation, as we arrive at a very large and negative entropy.

\begin{figure}[h!]
    \centering
    \includegraphics[width=0.8\textwidth]{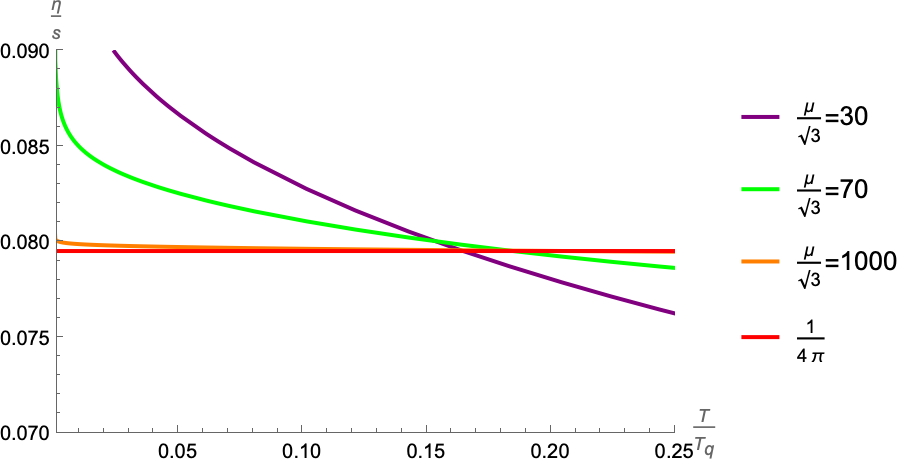}
    \caption{Temperature dependence of $\eta/s$ in fixed $C=1$ and different $\mu$.}
    \label{eta/s rh}
\end{figure}

There are various scales in the problem, and we can subsequently construct various dimensionless ratios. We have already chosen the regime where $T<T_q<r_h$. Let us view the various regions in terms of the dimensionless quantities $r_h/T_q$ and $T/T_q$.  If we take $r_h/T_q \to \infty$ first, the result of $\eta/s$ remains strictly equal to the $\frac{1}{4\pi}$ value. This is the regime explored in various previous works \cite{Edalati:2009bi, Moitra:2020dal}. Here, we clarify that an underlying requirement is to keep the temperature finite to make the $\log(T/T_q)$ term in the denominator in Eq.~\eqref{eta/s expansion} sub-leading. This behavior can be seen in Fig.~\ref{eta/s rh}, where when $\mu/T_q$ is very large, the curve of $\eta/s$ gradually coincides with the $1/4\pi$ results. A relevant discussion regarding the stabilizing role of large chemical potentials in hydrodynamics was presented in \cite{Davison:2013bxa}. Again, the intuition is correct, granted one keeps a small but non-zero temperature, rather than taking the temperature all the way to zero first. For smaller $\mu/T_q$, quantum corrections are significant and the value of $\eta/s$  deviates from $1/4\pi$. Recently, quantum corrections to $\eta/s$ have also been evaluated in various detailed studies with different approaches \cite{PandoZayas:2025snm,QuantumEtaOverS}. Our results reported here are compatible with those works within the range of validity of our approximation. In particular, given that our approach captures the one-loop entropy correctly, we believe this is the dominant effect responsible for the quantum curve of $\eta/s$ crossing $(1/4\pi)$ in Fig. \ref{eta/s 1/4pi}.

For the first order charge current, we have
\begin{equation}
    J_{\nu}^{(1)}=\frac{1}{4\pi G}e^{-i\omega v + ik_{x}x} \bigg(0,\, J_{x}^{(1)},\, J_{y}^{(1)}\bigg)\, ,
\end{equation}
where  
\begin{equation}
    \begin{aligned}
        J_{x}^{(1)} & = -i\sqrt{3}(k_x \delta r_h -r_h \omega \delta \beta_x) -\frac{ik_x \delta r_h }{r_h}\sqrt{\frac{T}{C}}+\frac{i T}{2\sqrt{3}r_h^2 C} \Bigg[ -58k_x \delta r_h\\
        & \quad + 21r_h \omega \delta \beta_x - 36k_x \log \left(\frac{3T}{C r_h^2}\right) \delta r_h + 9r_h \omega \log \left(\frac{3T}{C r_h^2}\right) \delta\beta_{x}\Bigg] + \mathcal{O}(T^{3/2})\, .
    \end{aligned}
\end{equation}
and 
\begin{equation}
    \begin{aligned}
        J_{y}^{(1)}&=i\sqrt{3}r_h \omega \delta \beta_y -\frac{i\sqrt{3}\omega \delta T\beta_y }{2 C r_h} \left(-7 +6\log(r_h) - 3\log \left(\frac{3T}{C}\right)\right) + \mathcal{O} (T^{2})\,.
    \end{aligned}
\end{equation}
Rewriting the charge current into its covariant form, we can obtain
\begin{equation}
    J_{\nu}=J_{\nu}^{(0)}+J_{\nu}^{(1)}=\rho u_{\nu}-\chi_{1}\mathfrak{a}_{\nu}-\chi_2P_{\nu}^{\lambda}\partial_{\lambda}\mu(x^{\sigma})\, .
\end{equation}
where $\mathfrak{a}_{\nu} = u^{\lambda}\partial_{\lambda}u_{\nu}$ is the acceleration, $P_{\mu\nu}=\eta_{\mu\nu}+u_{\mu}u_{\nu}$ is the projection tensor, and $\mu(x^{\sigma})=\sqrt{3}r_h +\sqrt{3}\delta r_h e^{-i\omega v +ik_x x}$. The transport coefficients are 
\begin{equation}
    \begin{aligned}
        \chi_{1} & = \frac{\sqrt{3}}{8\pi G Cr_h } \left(2C r_h^2 +7T +3 T\log \left(\frac{3T}{C r_h^2}\right)\right),
        \\ \chi_{2} & = \frac{1}{12\pi G C r_h^2} \left[\sqrt{3} r_h \sqrt{C T} + \left(3C r_h^2 +29 T +18 T \log \left(\frac{T}{C r_h^2}\right) \right)\right]\, ,
    \end{aligned}
\end{equation}
where $\chi_1$ is the charge diffusive coefficient associated with the gradient of fluid velocity, and $\chi_{2}$ is the thermal diffusivity associated with the gradient of chemical potential. The temperature-dependent terms, such as  $T\log T$ in the transport coefficients, are from quantum corrections.
\begin{figure}[h]
    \centering
    \includegraphics[width=0.48\textwidth]{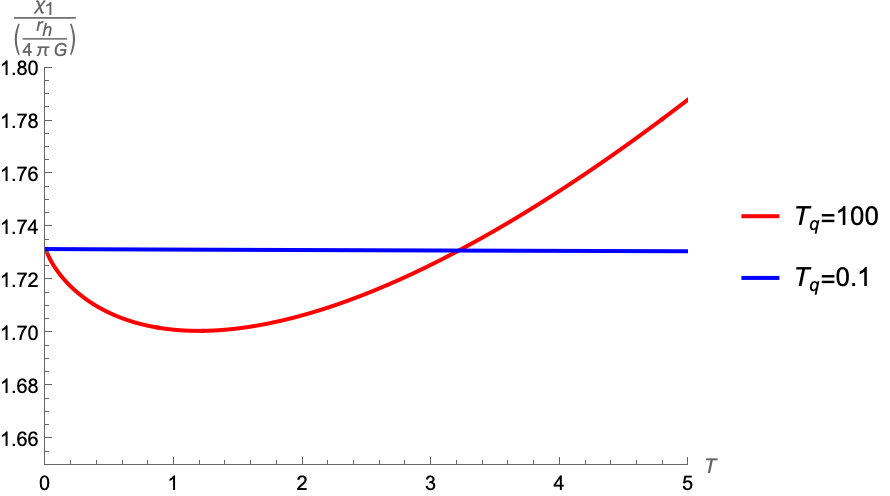}
    \includegraphics[width=0.48\textwidth]{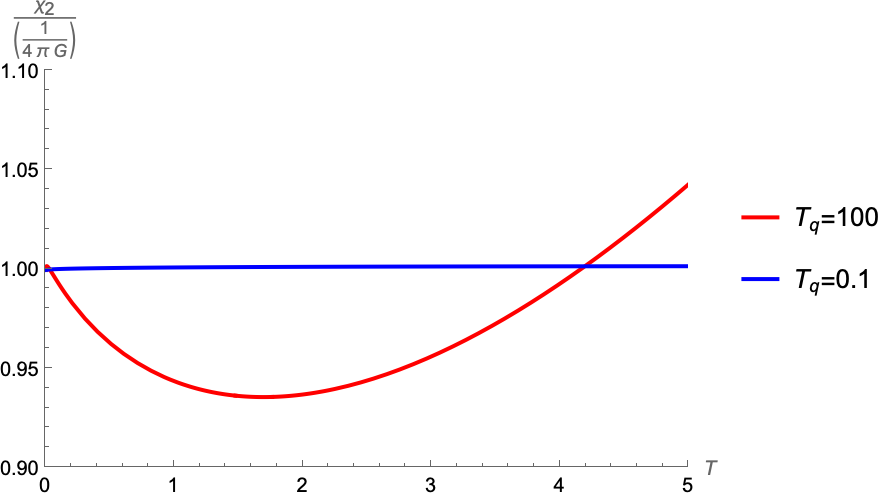}
    \caption{Temperature dependence of $\chi_1$ and $\chi_2$ at $r_h=100$, where the left figure is $\chi_{1}/(\frac{r_h}{4\pi G})$ and the right figure is $\chi_2/(\frac{1}{4\pi G})$. }
    \label{charge current coefficients}
\end{figure}

Keeping only the classical terms above, we recover results similar to those presented in  \cite{Moitra:2020dal}. Namely, $\chi_{1}=\frac{\sqrt{3}r_h}{4\pi G}, \chi_{2}=\frac{1}{4\pi G}$. As shown in  Fig.~\ref{charge current coefficients}, when quantum correction is leading ($T<T_q$), the temperature dependence of transport coefficients will be highly corrected, and when $T>T_{q}$, the value of $\chi_1, \chi_2$ is the same as classical extremal results in\cite{Moitra:2020dal}.

\subsection{Dispersion Relation}

We can now calculate the dispersion relation of the corresponding effective hydrodynamic modes. By using the conservation equations of the stress tensor and the charge current 
\begin{equation}
    \partial_{\mu\nu}T^{\mu\nu}=0,\quad \partial_{\mu}J^{\mu}=0\, ,
\end{equation}
and by keeping the zeroth and first orders of the stress-energy tensor and charge current to linear order in $T$, we can obtain four equations, which are explicitly given in App.~\ref{App:Pert-Details}.

We summarize those four kinds of hydrodynamic modes  in coordinate-free forms as

\begin{equation}
    \begin{aligned}
        \omega & = -i\gamma |\vec{k}|^2=\frac{-i|\vec{k}|^2}{18T_q^{-1} r_h^3(3T+2C(r_h^2+2\pi^2 T))}\Big(-90 r_h^2 TT_q^{-1}+4\sqrt{3}T_q^{-5/2}\pi^2 r_h^3 T^{3/2}-594T^2\\
        & \quad + 3T_q^2 (r_h^4 -40 \pi^2 r_h^2 T^2)+6\sqrt{3}r_h(r_h^2 \sqrt{T_{q}^{-3}T}+24\sqrt{T_q^{-1} T^3})\Big)\\
        & \text{(Shear mode)}\\
        \omega & =\pm v_s |\vec{k}| -i\Gamma |\vec{k}|^2=\pm\frac{|\vec{k}|}{\sqrt{2}}-\frac{i|\vec{k}|^2T_q}{36T_q^{-1}r_h^3 (3T+2T_{q}^{-1}(r_h^2 +2\pi^2 T^2))}\Big(-90T_q^{-1} r_h^2 T \\& 
        \quad + 4\sqrt{3}T_q^{-5/2}\pi^2 r_h^3 T^{3/2}-594T^2+3T_q^{-2}(r_h^4 -40\pi^2 r_h^2 T^2)+6\sqrt{3}r_h(r_h^2 \sqrt{T_q^{-3}T}+24 \sqrt{T_q^{-1}T^3})\Big)\\
         & \text{(Sound mode)}\\
          \omega & = -iD |\vec{k}|^2=-\frac{i|\vec{k}|^2T_q^2}{18 \sqrt{3}r_h^5}\Bigg[9\sqrt{3}T_{q}^{-2}r_h^4+3T_q^{-1}r_h^2 (29\sqrt{3}T +3r_h\sqrt{T_qT})-T(25\sqrt{3}T+222r_h\sqrt{T_q^{-1}T})\\
    & \quad - 9T(-6\sqrt{3}T_q^{-1}r_h^2 +20\sqrt{3}T+27r_h\sqrt{T_q^{-1} T}) \log \left(\frac{3TT_q}{r_h^2}\right) - 162\sqrt{3}T^2 \log \left(\frac{3TT_q}{ r_h^2}\right)^2\Bigg]
    \\
    &\text{(Charge diffusion mode)}
    \end{aligned}
\end{equation}
We observe the same four kinds of hydrodynamic modes as in the classical case; the only difference is that the coefficients in the dispersion relation have been corrected. The extra Schwarzian mode in NHR can be identified as an energy fluctuation mode as first described in \cite{Jensen:2016pah}. A framework discussed in \cite{Arean:2020eus} formulates the hydrodynamic problem in terms of Green's function. We now identify the emergent infrared mode there with the infrared mode that exists in the $AdS_2$ NHR region. We have tracked how that mode affects the asymptotically AdS$_4$ hydrodynamics dispersion relations.

\begin{figure}[h!]
    \centering
    \includegraphics[width=0.45\textwidth]{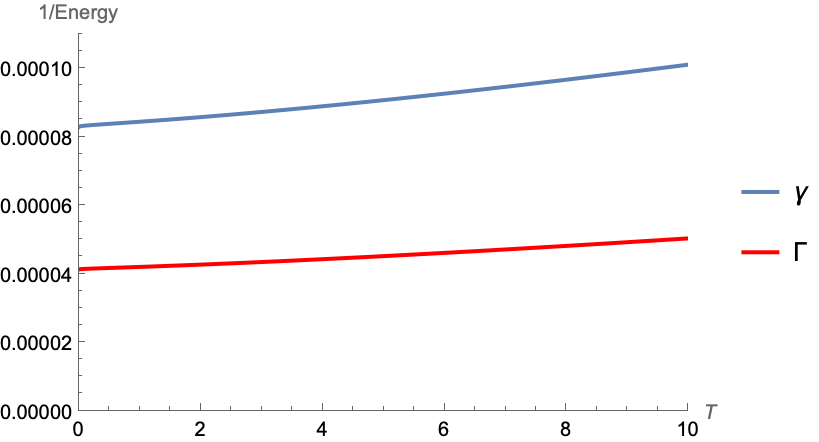}
    \includegraphics[width=0.45\textwidth]{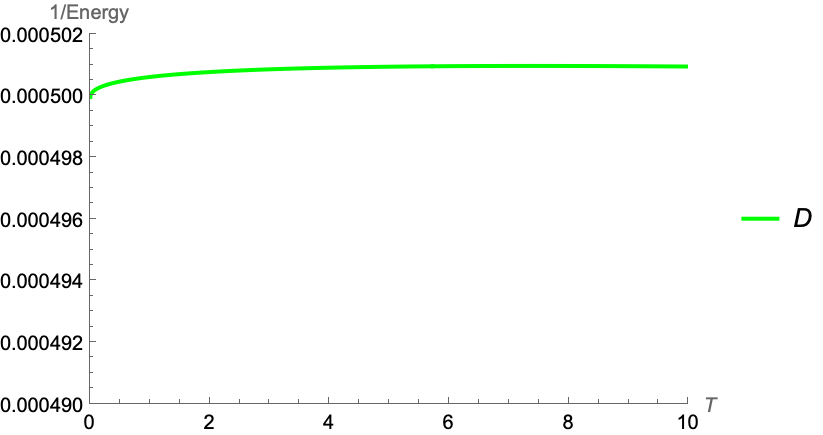}
    \caption{Dispersion relation coefficients at $C=0.1\,, r_h=1000$.}\label{Plot:DispersionCoeff}
\end{figure}

We have worked in the approximation where these emergent Schwarzian modes can be quantized and treated as a quantum-averaged contribution to the original long-wavelength hydrodynamic modes.  We illustrate the behavior of the coefficients in the dispersion relations in Fig.~\ref{Plot:DispersionCoeff} 

It appears that the dissipative coefficients in each mode will remain stable as the temperature approaches zero, which is expected due to the presence of an event horizon in the zero-temperature limit. In the low temperature regime, the  coefficients behave as 

\begin{equation}
    \begin{aligned}
        \gamma & = \frac{1}{12r_h}+\frac{\sqrt{TT_q}}{2\sqrt{3}r_h^2}-\frac{21TT_q}{8r_h^3}+\mathcal{O}(T^{3/2})\\
        \Gamma & = \frac{1}{24r_h}+\frac{\sqrt{TT_q}}{4\sqrt{3}r_h^2}-\frac{21TT^{q}}{16 r_h^3}+\mathcal{O}(T^{3/2})\\
        D & = \frac{1}{2r_h}+\frac{\sqrt{TT_q}}{2\sqrt{3}r_h^2}+\frac{\left(29+18\log(\frac{3TT_q}{r_h^2})\right) T T_q}{6r_h^3}+\mathcal{O}(T^{3/2})\,.
    \end{aligned}
\end{equation}
The leading order of these coefficients is the same as \cite{Moitra:2020dal}, and quantum corrections yield only sub-leading $\sqrt{T}$ contributions. Given that quantum corrections are central to changing the thermodynamic behavior, one might have expected that the leading behavior in the dispersion coefficients would also be affected. This result requires deeper investigation and should be considered in other implementations of the quantum corrections.

\section{Summary and Discussion}\label{sec:Discussion}

In this paper, we revisit the fluid/gravity correspondence in the regime of very low temperatures. We implemented a general method for introducing quantum corrections at the level of the equations of motion, in which we quantum-average the emergent infrared degrees of freedom as an effective contribution to the original long-wavelength hydrodynamics modes. Using insights developed in the quantum treatment of JT gravity, we focused on treating the breaking of time rescaling symmetry of the near-horizon geometry. 

We first reviewed previous analyses and highlighted the ubiquitous presence of a non-analytic term, $\log(\omega)$, in the holographic fluid derivative expansion when the temperature is taken to zero first. We attribute the presence of such logarithmic terms to the neglect of issues related to time rescaling symmetry and its breakdown in the near-horizon region. Indeed, turning on a temperature explicitly breaks the AdS$_2$ rescaling symmetry. Including the Schwarzian modes in the near-horizon region also affects rescaling symmetry. We then studied various temperature regimes in comparison with the new scale of quantum fluctuations, $C^{-1}=T_q$, which represents the energy scale of Schwarzian fluctuations.

By comparing different regime choices, we showed that the appearance of $\log(\omega)$ is typical of a particular order of limits, which leads to a non-analytic derivative expansion in the fluid description. The guiding principle is that a well-defined hydrodynamic description requires the system to be in local equilibrium. Consequently, if the linearized perturbation scale, $\omega$, is greater than the characteristic energy scale of the system, a single perturbation will destroy the equilibrium configuration manifested in the appearance of $\log(\omega)$ terms. Moreover, we surprisingly found that in the quantum regime $T<T_q$, quantum fluctuations enhance the relaxation process and help stabilize the fluid description at low $T$.

We calculated the quantum-corrected first-order fluid stress tensor and charge current using the bulk in the fluid/gravity correspondence. We evaluated $\eta/s$ and the results show that once the temperature is finite and near the scale of $\frac{1}{C}$, the value of $\eta/s$ might be lower than the $1/4\pi$ bound. With the results of the first-order constitutive relations for the fluid, we calculated the dispersion relations for four hydrodynamic modes. The emerging picture is one in which the quantum fluctuations of the Schwarzian modes, which were treated hydrodynamically in \cite{Jensen:2016pah}, can be considered as averaged to lead to an effective contribution to the original long-wavelength modes. We also checked the extremal limit of diffusion coefficients of these modes, and the results show that the diffusive coefficients are not affected at leading order in a small temperature expansion.

\subsection{Low Temperature Fluids and  Quantum Black Holes}
The fluid/gravity correspondence includes a deep connection between dissipative processes in fluids and the physics of black holes. Since, in this work, we have modified this correspondence by including certain quantum corrections, it is only natural to discuss their implications for quantum dissipation and quantum absorption in black holes.

Recall that in the semi-classical regime, it has been shown that the shear viscosity is proportional to the absorption cross section \cite{Policastro:2001yc}
\begin{equation}
   \eta = \lim_{\omega\to 0}\frac{1}{2\kappa^2}\sigma_{abs}(\omega)=\frac{1}{2\kappa^2}\sigma(0)\,.
\end{equation}
It is a classical result in general relativity that the absorption cross section is proportional to the area of the black hole horizon \cite{Das:1996we}. Furthermore, the area of the black hole horizon is the entropy of the black hole. 

\begin{figure}[h]
    \centering
    \includegraphics[width=0.75\textwidth]{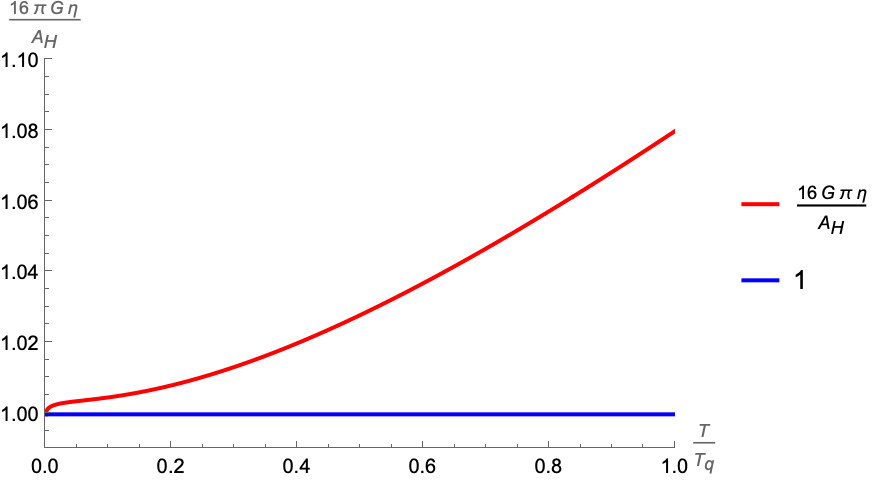} 
    \caption{Ratio of quantum-corrected $\eta$ and horizon area at $C=1,\,r_h=100$}
    \label{fig: ratio of eta}
\end{figure}

Recent developments have implications for these relations, as they are now affected in the presence of quantum corrections generated by the near-horizon fluctuations. In the quantum regime of $T<T_q$, the dynamics of near-extremal near-horizon $AdS_2$ throat becomes leading, and the absorption cross section is highly enhanced \cite{Emparan:2025sao} (see also \cite{Biggs:2025nzs}). Those results imply that when the energy of the black hole is very low $E_{BH}\ll T_{q}$, the absorption cross-section increases dramatically,
\begin{equation}
    \sigma_{abs}(\omega\to 0)\approx A_{H}\frac{1}{\pi}\sqrt{\frac{T_{b}}{2E_{BH}}}> A_{H}\,.
\end{equation}
Our analysis, plotted in Fig.~\ref{class-quant eta}, is compatible with such an observation.  Our calculation of $\eta$ also supports the observation that quantum corrections enhance the absorption process of black holes away from the area law; it can also be seen in Fig.~\ref{fig: ratio of eta} clearly. \\
\\
We need to emphasize that the validity of our approach remains when $T<T_q$ and $T$ is finite. For  $T$ very close to $T_0$ in Eq. \eqref{Eq:T0}, the approximation breaks down and contributions from other topologies might become important \cite{Antonini:2025nir}.

It was recently pointed out that the standard connection between semiclassical absorption cross section, area, and entropy of black holes should be revisited in the presence of quantum corrections \cite{Emparan:2025sao}. The energy distribution is arguably a more appropriate proxy 
\begin{equation}
    \sigma_{abs}>A_H >4G \log \rho(E)\,. 
\end{equation}
We expect the quantum value of $\eta/s$ to be greater than $1/4\pi$ in the very low temperature regime. This intuition is borne out in Fig.~\ref{eta/s 1/4pi}.

Let us proceed with some remarks that we believe connect the quantum aspects of black holes to very low-temperature fluid dynamics.

\begin{enumerate}
    \item \textbf{Stronger absorption rate ensures fast relaxation towards equilibrium:}\par
    In Sec.~\ref{sec:Massless Scalar Model}, we have discussed the issue of the existence of $\log(\omega)$ and proposed two appropriate regime choices where the fluid can achieve local equilibrium. Alternatively, one can think of the emergence of $\log(\omega)$ terms as a signal of the breakdown of the hydrodynamics derivative expansion, whereby the derivative expansion does not converge.

    $\quad$ In the semi-classical limit ($T_q < T$), we can choose $T_q<\omega <T$ to make sure the external perturbation scale $\omega$ is smaller than the black hole energy, and the near-equilibrium dissipative process can relax the system back to local equilibrium.

    $\quad$ For the quantum regime ($T < T_q$), the mechanism for relaxation to equilibrium is quite different. We find that, by choosing $T<\omega <T_q$, as long as the external perturbation scale is smaller than $T_q$, the derivative expansion can surprisingly converge, even for $T<\omega$.
    
    $\quad$ One possible interpretation, using the notion of quantum absorption, $\sigma_{abs}$, suggests that quantum corrections significantly enhance the energy absorption process for black holes. This process ultimately strengthens the relaxation of any perturbation with $\omega<T_q$, and can quickly bring the perturbed system back to equilibrium. 

    \item \textbf{Interaction between quantum IR modes and classical external source:}

    The naive symmetries of the near-horizon $AdS_2$ throat lead to both the IR $\log(\omega)$ non-local terms and to extra IR mode issues that break the hydrodynamics description. Semi-classically, it has been studied that the $SL(2,\mathbb{R})$ symmetry of the near-horizon $AdS_2$ geometry generates new poles in the IR Green's function, indicating the existence of extra IR modes \cite{Arean:2020eus, Liu:2021qmt}. In this framework, the collision between gapless hydro-mode poles and these IR mode poles signals the breakdown of the hydrodynamics description.
    
    $\quad$ In recent works \cite{Emparan:2025sao, Biggs:2025nzs}, coupling an external classical matter to the effective Schwarzian action in the $AdS_2$ throat incorporates quantum effects from the Schwarzian mode in the physics under consideration. The quantum transition rate is altered, leading to situations where the energy of external sources is rapidly absorbed into a quantum black hole. One can view both these processes as dual ways of probing the quantum black hole.
    
     $\quad$ Our main approach considered classical tensor and vector field perturbations in the presence of quantum fluctuations by introducing a quantum average in the EoMs. Both approaches consider quantum Schwarzian corrections and lead to a similar conclusion: the interaction between the IR mode and hydro-modes is very important, especially in the quantum regime, $T<T_q$. 
     
     $\quad$ Classically, the existence of IR modes seems very likely to break the fluid description, but surprisingly, after quantization and averaging, they play a crucial role in stabilizing the quantum black hole and quantum fluid. A similar consideration regarding the construction of an effective action for the IR mode is given by \cite{Liu:2024tqe}.

    \item \textbf{Toward  quantum fluids:}
    
    The concept of mean free path and mean free time of particles is central to establishing standard notions of hydrodynamics. The perils of extending these notions straightforwardly to the quantum hydrodynamics regime were recently highlighted in  \cite{Emparan:2025sao}. In particular, the need to confront a certain sub-Planckian regime was put forward.

    $\quad$ When we compute the shear viscosity, we use the fluctuation-dissipation theorem (Green-Kubo formula). However, the formula is primarily used for classical fluid definitions, and in the quantum case, one should extend the regime of the fluctuation-dissipation theorem to include aspects of the quantum regime.
    
 \end{enumerate}

\subsection{Outlook for Future Directions}

Let us briefly outline some possibilities for extending the approach of the quantum-corrected EoM and examining the quantum behavior of a low-temperature fluid.

\begin{itemize}
   
    \item Topological fluid excitation:\\
    One interesting feature of JT/Schwarzian is that the partition function, in principle, contains contributions from saddles of different topology. Recent discussions in a related context have appeared in \cite{Antonini:2025nir}, and it would be interesting to explore similar corrections in the Fluid/Gravity correspondence.
    
\item Graviational path integral:\\
It would be quite interesting to approach the problem of including quantum corrections directly from the four-dimensional gravitational path integral point of view. 

\end{itemize}
\section*{Acknowledgements}
We would like to thank Yanyan Bu, Blaise Gout\'eraux, Kristan Jensen, Xiao-Long Liu, Thomas Mertens, Upamanyu Moitra, Cheng Peng, Konraad Schalm, Kostas Skenderis, Weijie Tian, and Jingchao Zhang for helpful discussions. This work is supported in part by the NSFC under grants No.~12375067, No.~12147103, and No.~12247103.  LAPZ is partially supported by the U.S. Department of Energy under grant DE-SC0007859. J.N. would like to thank the International Centre for Theoretical Physics (ICTP) and the Niels Bohr Institute (NBI) for their warm hospitality during the final stage of this work. LAPZ thanks the Beijing Institute for Mathematical Sciences and Applications, ICTP-AP of the Chinese Academy of Sciences, and Peng Huanwu Center for Fundamental Theory at USTC, Hefei, for hospitality at various stages of this work.

\newpage
\appendix



\section{Perturbation Theory: Details}\label{App:Pert-Details}

 In this appendix, we present the explicit expression for the first-order linearized perturbation solution and provide details on computing the dispersion relations. The method we used to compute these is given in Sec.~\ref{sec:Near-Ext Fluid Gravity Correspondence}. The zeroth-order perturbation can be obtained by firstly substituting Eq.(\ref{adding coordinates dependence in fluid quantities}) into the background metric and then by subtracting the original background to obtain the perturbation parts. Therefore, we can get the zeroth-order perturbations as 
\begin{equation}
    \begin{aligned}
        h_{vv}^{(0)}&=\frac{2}{C r^2}\Big((6C r_h^2 (r_h -r)+3(r+3r_h)T +4\pi^2 C(r+3r_h)T^2 )\delta r_h\\
        &+(r^2 +2r r_h +3 r_h^2)(3+8\pi^2 CT)\delta T\Big)\\
        h_{vi}^{(0)}&=\frac{1}{C r^2}\Big(Cr_h^3(3r_h -4 r)+3(r^2 +2rr_h +3 r_h^3)T +4 C \pi^2 (r^2 +2 r r_h +3 r_h^2)T^2 \delta \beta_{i}\Big)\\
        h_{r,i}^{(0)}&=-\delta \beta_{i}\\
        a_{v}^{(0)}&=-\frac{2\sqrt{3}r_h\delta r_h}{r}\\
        a_{i}^{(0)}&=-g(r)\delta \beta_{i}\,.
    \end{aligned}
\end{equation} \\
Those zeroth-order perturbations play the role of source terms in the first-order EoMs; in the later subsection, we will calculate the first-order perturbation based on these sources.

\subsection{The First-Order Perturbation}

As we have explained in the main text, the metric and gauge field components have been classified into tensor, vector, and scalar sectors; we will now investigate them separately.

\subsubsection{Tensor Sector}

In the tensor sector we have $(g_{xx}^{(n)} - g_{yy}^{(n)}) / 2$ and $g_{xy}^{(n)}$. The classical EoM for the tensor sector has a form similar to that of the massless scalar. For example, the classical EoM for the $g_{xy}$ component is
\begin{equation}
    -\frac{1}{2} \Big[r^4 f(r)\, \big(h^{y}_{x} (r) \big)'\Big]' + \frac{i}{2} \Big[2\omega r \big(r\, h^y_{x}(r) \big)' + k_{x} \big(r^4 f(r) h^y_{r} (r) + r^2 h^{y}_{v}(r) \big)'\Big] + \frac{1}{2} k_{x}\omega r^2 {h^{y}}_{r}(r) = 0\, .
\end{equation}
In the first-order calculation, we can ignore the second-order $\omega, k_{x}$ term and the derivative term with $\omega$. Consequently, the quantum-corrected EoM can be simplified to 
\begin{equation}
    \Big[r^4 f(r)Q_{cor}(r) \big(h^{y\, (1)}_x (r)\big)'\Big]' = i \Big[k_{x} \big(r^4 f(r)Q_{cor}(r) h^{y\, (0)}_r + r^2 h^{y\, (0)}_v\big)' \Big]\, .
\end{equation}
Using the non-corrected value for the zeroth-order source, we obtain
\begin{equation}
\begin{aligned}
    s_{xy}^{(0)} &= i \Big[k_{x} \big(r^4 f(r)Q_{cor}{{h^y}_{r}^{(0)}} + r^2 {{h^{y}}_{v}^{(0)}}\big)'\Big] \\&= -\frac{2ik_x}{C r^3}\Big(6r_h (r+3r_h)T +C(r^4 +8\pi^2 r_h (r+3r_h)T^2)\delta \beta_{x}\Big)\, .
\end{aligned}
\end{equation}
Then, the quantum-corrected EoM in the FAR can be written as
\begin{equation}
    \frac{d}{dr} \Big(r^2 F(r)_{FAR}\, Q_{cor}(r) \frac{d}{dr} (r^{-2} h_{yx,\, cor}^{(1)})\Big) = s_{xy}^{(0)}\, .
\end{equation}
As we can see, the only difference between the classical and the quantum-corrected EoMs is the additional quantum correction factor $Q_{cor}(r)$. Therefore, the solution can be directly written down as
\begin{equation}
    r^{-2}h_{yx}^{(1)} = c_{yx,2}^{(1)} - \int_{r}^{\infty} \frac{dr'}{r'^2 F(r')_{FAR}\, Q_{cor}(r)} \left(c_{yx,1}^{(1)}+\int_{r_{B}}^{r'}dr'' s_{xy}^{(0)}\right)\, ,
\end{equation}
where we can fix $c_{yx,2}^{(1)} = 0$ using the $r \to \infty$ normalizable condition, and $c_{yx,1}^{(1)}$ can be fixed by matching with the NHR solution, just like the scalar model case. As we analyzed in the massless scalar case, matching the inner solution and outer solution is equivalent to extending the outer solution's integration limit from $r_{B}$ to $r_{+}$. Therefore, We can directly set $c_{yx,1}^{(1)} = 0$ and change the inner integration limit from $r_{B}$ to $r_{h}$. Then, we can obtain
\begin{equation}
    r^{-2}h_{yx}^{(1)} = 2ik_{x}\delta \beta_{y}\int_{r}^{\infty}\frac{dr'}{r'^2F(r')_{FAR}Q_{cor}(r)}\int_{r_{h}}^{r'}dr'' s_{xy}^{(0)}\, .
\end{equation}
The explicit expression of $h_{yx}^{(1)}$ 
is

\begin{equation}\label{Eq:App-h-yx}
    \begin{aligned}
        h_{yx}^{(1)} & = -\frac{ik_x r^2 \delta \beta_{y}}{216 C^2 r_h^5}\Bigg (-18\sqrt{2}\pi C^2 r_h^4 +4\sqrt{6}\pi^3 r_h^3 C^{5/2}T^{3/2}-479\sqrt{2}\pi T^2 -\frac{162r_h^4 T^2}{(r-r_h)^4}\\
        & \quad + 6\sqrt{6}\pi r_h^3 \sqrt{C^3 T}+137\sqrt{6}\pi r_h\sqrt{C T^3}-38\sqrt{2}C\pi r_h^2 T (3+4C \pi^2 T)+\frac{72 r_h^3 T(2T +\sqrt{3}r_h \sqrt{CT})}{(r-r_h)^3}\\
        & \quad + \frac{432 r_h T (9T+Cr_h^2 (3+4C \pi^2 T))}{r}+\frac{12 r_h}{r-r_h}\Big (4\sqrt{3} C^{5/2}\pi^2 r_h^3 T^{3/2}+292T^2 +6\sqrt{3}r_h^3 \sqrt{C^3 T}\\
        & \quad + 149 \sqrt{3}r_h\sqrt{C T^3} +4Cr_h^2 T (3+4C \pi^2 T)\Big)-\frac{18r_h^2T}{(r-r_h)^2}\Big(149T +2r_h(2\sqrt{3}\sqrt{CT}\\
        & \quad + Cr_h (3+4C\pi^2 T)) \Big)-2\sqrt{2}\Big(-18 C^2 r_h^4 +4\sqrt{3}C^{5/2}\pi^2 r_h^3 T^{3/2} -479 T^2+6\sqrt{3}r_h^3 \sqrt{C^3 T}\\
        & \quad + 137 \sqrt{3}r_h \sqrt{C T^3}-38 C r_h^2 T(3+4C \pi^2 T)\Big)\mathrm{arccot}\Big(\frac{\sqrt{2}r_h}{r+r_h}\Big)-864 T (18 T \\
        & \quad + C r_h^2 (3+4C \pi^2 T))\log(r)+2\Big(36 C^2 r_h^4 +16\sqrt{3}C^{5/2}\pi^2 r_h^3 T^{3/2} +5257 T^2 +24 \sqrt{3}r_h^3\sqrt{C^3 T}\\
        & \quad + 584 \sqrt{3}r_h\sqrt{C T^3}+298 C r_h^2 T(3+4C \pi^2 T)\Big)\log(r-r_h)-\Big( 36 C^2 r_h^4 +16\sqrt{3}C^{5/2}\pi^2 r_h^3 T^{3/2}\\
        & \quad - 2519 T^2 +24\sqrt{3}r_h^3 \sqrt{C^3 T}+584 \sqrt{3}r_h\sqrt{C T^3}-134 C r_h^2 T (3+4 C \pi^2 T)\Big)\log(r^2 +2r r_h +3 r_h^2)\Bigg)\,.
    \end{aligned}
\end{equation}
As we can see in Equation \eqref{Eq:App-h-yx}, the explicit expression (up to $\mathcal{O}(T^2)$) is quite lengthy, and for the vector sector, it is even lengthier than this expression. Therefore, from now on, we will only show the asymptotic expression at $r\to \infty$ for all first-order solutions, and we have\footnote{Note that here the results have been taken from the expansion with respect to the quantum correction regime ($T<T_q$).} If one takes the semiclassical regime expansion ($T_q < T$), one can finally obtain the semiclassical area law of $\eta$.
\begin{equation}
\begin{aligned}
        h_{yx}^{(1)}\Big |_{r\to \infty} & = ik_x \delta\beta_{y}r-\frac{ik_x}{9C^2 r_h^2 r}\Big(-90C r_h^2 T +4\sqrt{3}C^{5/2}\pi^2 r_h^3 T^{3/2}-594 T^2 +3C^2 (r_h^4-40 \pi^2 r_h^2 T^2)\\
        & \quad + 6\sqrt{3}r_h (r_h^2 \sqrt{C^3 T}+24 \sqrt{C T^3})\Big)\delta \beta_y+\frac{ik_x r_h(-3T +2 C(r_h^2 -2\pi^2 T^2))\delta \beta_y}{2Cr^2 }+\mathcal{O}(\frac{1}{r^3})\,.
\end{aligned}
\end{equation}
Similarly, we obtain
\begin{equation}
\begin{aligned}
        r^2 \alpha^{(1)}\Big |_{r\to \infty} & = \frac{1}{2}(h_{xx}^{(1)}-h_{yy}^{(1)})\Big |_{r\to \infty}\\
        & = ik_x \delta\beta_{x}r-\frac{ik_x}{9C^2 r_h^2 r}\Big(-90C r_h^2 T +4\sqrt{3}C^{5/2}\pi^2 r_h^3 T^{3/2}-594 T^2 +3C^2 (r_h^4-40 \pi^2 r_h^2 T^2)\\
        & \quad + 6\sqrt{3}r_h (r_h^2 \sqrt{C^3 T}+24 \sqrt{C T^3})\Big)\delta \beta_x+\frac{ik_x r_h(-3T +2 C(r_h^2 -2\pi^2 T^2))\delta \beta_x}{2Cr^2 }+\mathcal{O}(\frac{1}{r^3})\,.
\end{aligned}
\end{equation}
Then $h_{xx}^{(1)}$ and $h_{yy}^{(1)}$ can be obtained from $\sigma^{(1)}(r)$ and $\alpha^{(1)}(r)$.

\subsubsection{Vector Sector}

In the vector sector, we have the components $g_{vi}^{(n)}$ and $A_{i}^{(n)}$, which obey coupled EoMs. The classical EoMs of $a_{x}$ and $h_{xv}$ can be written as \cite{Moitra:2020dal}:
\begin{align}
    \Big[\frac{1}{r^2} \big(r^2 f(r)a'_{x}(r)\big)' + g'(r) {h^x}_{v}'(r)\Big] - \frac{i}{r^2} \Big[k_{x}a'_{v}(r) + 2\omega a'_{x}(r) - \omega r^2 g'(r) {h^{x}}_{r}(r)\Big] & = 0\, ,\\
    \Big[\frac{1}{2r^2} \big(r^4 {h^x}_{v}'(r)\big)' + 2g'(r)a_{x}'(r)\Big] + \frac{i}{2} \Big[\frac{\omega}{r^2}(r^4 {h^x}_{r})' - k_{x}r^2 \left(\frac{h_{vr}}{r^2}\right)' + k_{x} (\alpha' - \sigma')\Big] & = 0\, .
\end{align}
The classical EoMs of $a_{y}$ and $h_{yv}$ are
\begin{align}
    \Big[\frac{1}{r^2} \big(r^2 f(r) a'_{y}\big) + g'(r)\, {h^{y}}_{v}'(r)\Big] - \frac{i\omega}{r^2} \Big[2a'_{y}(r) - r^2 g'(r){h^{y}}_{r}(r) \Big] - \frac{k_x^2 a_{y}(r)}{r^4} = 0\, ,\\
    \Big[\frac{1}{2r^2} \big(r^4 {h^y}_{v}'(r) \big)' + 2g'(r)\, a_{y}'(r)\Big] + \frac{i}{2} \Big[\frac{\omega}{r^2}(r^4 {h^y}_{r})' + k_{x} {h^y}_{x}'(r)\Big] + \frac{k_{x}^2}{2}{h^y}_{r}(r) = 0\, .
\end{align}
These equations take the general form of 
\begin{align}
    \frac{\partial}{\partial r} \left(r^2 f(r) \frac{\partial}{\partial r} A_{i}^{(n)}\right) + Q\frac{\partial}{\partial r} \left(r^{-2} g_{iv}^{(n)}\right) & = s_{i,1}^{(n)}\, ,\\
    \frac{\partial}{\partial r} \left(r^4 \frac{\partial}{\partial r} (r^{-2}g_{iv}^{(n)})\right) + 4Q\frac{\partial}{\partial r} A_{i}^{(n)} & = s_{i,2}^{(n)}\, ,
\end{align}
where $i\in \{x,y\}$ labels different sets of EoMs. By simply integrating the above equations once, we can obtain
\begin{align}
    r^2 f(r)\frac{\partial}{\partial r}A_{i}^{(n)} + Qr^{-2}g_{iv}^{(n)} & = c_{i,1}^{(n)} + \int^{r}_{r_{B}} dr' s_{i,1}^{(n)}(r')\, ,\\
    r^4 \frac{\partial}{\partial r}(r^{-2}g_{iv}^{(n)}) + 4QA_{i}^{(n)} & = c_{i,2}^{(n)} + \int_{r_{B}}^{r} dr' s_{i,2}^{(n)}(r')\, .
\end{align}
Using these equations, we can get 
\begin{equation}\label{eq:decoupled diff eq}
    \begin{aligned}
    & \frac{\partial}{\partial r}A_{i}^{(n)} = \frac{c_{i,1}^{(n)}}{r^2 f(r)} + \frac{1}{r^2 f(r)}\int_{r_{B}}^{r} dr' s_{i,1}^{(n)}(r') - \frac{Q}{r^4f(r)}g_{iv}^{(n)}\, ,
    \\ & \frac{\partial}{\partial r} \left(r^4 \frac{\partial}{\partial r} (r^{-2}g_{iv}^{(n)})\right) - \frac{4Q^2}{r^2f(r)}(r^{-2}g_{iv}^{(n)}) = s_{i,2}^{(n)} - \frac{4Q}{r^2f(r)} \left(c_{i,1}^{(n)} + \int_{r_{B}}^{r}dr's_{i,1}^{(n)}(r')\right)\, ,
    \end{aligned}
\end{equation}
where the second equation is a decoupled second-order differential equation for $g_{iv}^{(n)}$.

It is convenient to make the following transformation:
\begin{equation}
    r^{-2}g_{iv}^{(n)} = f(r)H_{i}^{(n)}\, .
\end{equation} 
Consequently,
\begin{equation}
    \begin{aligned}
        \frac{\partial}{\partial r} \left(r^4 f(r)^2 \frac{\partial H_{i}^{(n)}}{\partial r}\right) & = 4r^3 \left(f^2(r) \frac{\partial H_{i}^{(n)}}{\partial r}\right) + r^4 \frac{\partial}{\partial r} \left(f^2(r)\frac{\partial H_{i}^{(n)}}{\partial r}\right)
        \\ & =4r^3 f^2(r) \frac{\partial H_{i}^{(n)}}{\partial r}+2r^4 f(r) \frac{\partial f}{\partial r}\frac{\partial H_{i}^{(n)}}{\partial r}+r^4 f^2(r)\frac{\partial ^2 H_{i}^{(n)}}{\partial r^2}\, ,
    \end{aligned}
\end{equation}
and 
\begin{equation}
    \begin{aligned}
            {} & f(r) \frac{\partial}{\partial r} \left(r^4 \frac{\partial}{\partial r} (f(r)H_{i}^{(n)})\right) - \frac{4Q^2}{r^2}f(r)H_{i}^{(n)} \\
            =\,\, & 4r^3 f(r)\frac{\partial f}{\partial r}H_{i}^{(n)}+4r^3 f^2(r) \frac{\partial H_{i}^{(n)}}{\partial r}+r^4 f(r) \frac{\partial ^2 f}{\partial r^2}H_{i}^{(n)}
            + 2r^4 f(r) \frac{\partial f}{\partial r}\frac{\partial H_{i}^{(n)}}{\partial r} + r^4 f^2(r)\frac{\partial ^2 H_{i}^{(n)}}{\partial r^2} \\
            {} & - \frac{4Q^2}{r^2}f(r)H_{i}^{(n)}\, .
    \end{aligned}
\end{equation}
Moreover,
\begin{equation}\label{eq: vectoor constraint}
    4r^3 f(r)\frac{\partial f}{\partial r}H_{i}^{(n)} + r^4 f(r) \frac{\partial^2 f}{\partial r^2}H_{i}^{(n)} - \frac{4Q^2}{r^2}f(r)H_{i}^{(n)} = 0\, ,
\end{equation}
which is a component of the background Einstein equation and can be proved using the black brane metric factor $f(r) = 1 - \frac{2GM}{r^3} + \frac{Q^2}{r^4}$. Thus, we can finally get a simple differential equation for $H_{i}^{(n)}$ as follows: 
\begin{equation}\label{eq:Diff Eq for H_i}
    \frac{\partial}{\partial r} \left(r^{4}f^2(r)\frac{\partial H_{i}^{(n)}}{\partial r}\right) = s_{i}^{(n)}\, ,
\end{equation}
where the source term is
\begin{equation}
    s_{i}^{(n)} = f(r)s_{i,2}^{(n)} - \frac{4Q}{r^2} \left(c_{i,1}^{(n)}+\int_{r_{B}}^{r}dr' s_{i,1}^{(n)}(r')\right)\, .
\end{equation}
The solution to \eqref{eq:Diff Eq for H_i} is given by
\begin{equation}
    H_{i}^{(n)} = -\int_{r}^{\infty}\frac{dr'}{r'^4f(r')^2} \left(c_{i,3}^{(n)} + \int_{r_{B}}^{r'}dr''s_{i}^{(n)}(r'')\right) + c_{i,4}^{(n)}\, ,
\end{equation}
where the metric perturbation can be determined by the $H_{i}^{(n)}$, while the gauge field perturbation $A_{i}^{(n)}$ can be determined by the metric perturbation. As we can see, for each choice of the index $i$, there are four integration constants, $c_{i,1}^{(n)}$, $c_{i,2}^{(n)}$, $c_{i,3}^{(n)}$, and $c_{i,4}^{(n)}$. Two of these four constants can be easily fixed by the $r\to \infty$ boundary conditions for the metric and the gauge fields. More explicitly, $g_{iv}^{(n)}$ fixes $c_{i,4}^{(n)} = 0$, while $A_{i}^{(n)}$ fixes $c_{i,2}^{(n)}$ in terms of $c_{i,1}^{(n)}$ and $c_{i,3}^{(n)}$. For the remaining constants $c_{i,1}^{(n)}$ and $c_{i,3}^{(n)}$, we need to impose the Landau frame condition to fix them\footnote{We should remain the integration constant to the boundary stress tensor at first and then consider Landau frame condition for the stress tensor to give constraint on the integration constant.}.

For the $n=1$ case, we have the source terms as follows:
\begin{equation}
    \begin{aligned}
        s_{x,1}^{(0)} & = i \Big[k_x (a_{v}^{(0)})'- \omega g'(r)\, h_{xr}^{(0)}\Big] = \frac{\sqrt{3}i}{r^2} \Big(2r_{h}k_{x}\delta r_{h}-r_{h}^2 \omega \delta \beta_{x}\Big)\, ,\\
        s_{x,2}^{(0)} & = -ir^2 \bigg[\frac{\omega}{r^2}(r^4 {h^x}_{r}^{(0)})' - k_{x}r^2 \Big(\frac{h_{vr}^{(0)}}{r^2}\Big)'\bigg] = 2ir\omega \delta \beta_{x}\, ,\\
        s_{y,1}^{(0)} & = -i\omega r^2  g'(r){h^{y}}_{r}^{(0)}=-\frac{i\sqrt{3}r_h^2 \omega  \delta  \beta_{y}}{r^2}\, ,\\
        s_{y,2}^{(0)} & = -ir^2 \Big[\frac{\omega}{r^2}(r^4 {h^{y}}_{r})' + k_{x}\, (h^{y\, (0)}_{x})'\Big] = 2ir\omega \delta \beta_{y}\, .
    \end{aligned}
\end{equation}
The quantum-corrected EoMs are equal to the quantum average classical EoMs as
\begin{equation}
\begin{aligned}
    \frac{\partial}{\partial r} \Big(r^2 f(r) Q_{cor}(r) \frac{\partial}{\partial r} A_{i}^{(1)}\Big) + Q\frac{\partial}{\partial r}(r^{-2} g_{iv}^{(1)}) & = s_{i,1}^{(0)}\, ,\\
    \frac{\partial}{\partial r} \Big(r^4 \frac{\partial}{\partial r} (r^{-2} g_{iv}^{(1)})\Big) + 4Q\frac{\partial}{\partial r}A_{i}^{(1)} & = s_{i,2}^{(0)}\, .
\end{aligned}
\end{equation}
By repeating the same steps as before, we can get the following equation, similar to the classical one \eqref{eq:decoupled diff eq}:
\begin{equation}
    \frac{\partial}{\partial r} \Big(r^4 \frac{\partial}{\partial r}(r^{-2} g_{iv}^{(1)})\Big) - \frac{4Q^2}{r^2f(r)Q_{cor}(r)}(r^{-2} g_{iv}^{(1)}) = s_{i,2}^{(0)} - \frac{4Q}{r^2f(r)Q_{cor}(r)} \Big(c_{i,1}^{(1)} + \int_{r_{B}}^{r}dr' s_{i,1}^{(0)}(r')\Big)\, .
\end{equation}
After making a new transformation
\begin{equation}
    r^{-2}g_{iv}^{(1)} = f(r)H_{i}^{(1)}\, ,
\end{equation}
and use the same constraint Eq.(\ref{eq: vectoor constraint}), we have
\begin{equation}\label{eq:Diff Eq for H_i with quantum corr}
    \frac{\partial}{\partial r} \left(r^4 f^2(r)\frac{\partial H_{i}^{(1)}}{\partial r}\right) + 
    \left(1 - \frac{1}{Q_{cor}(r)}\right) \frac{4Q^2}{r^2}f(r)H_{i}^{(1)} = s_{i}^{(0)}\, .
\end{equation}
However, it is still not easy to solve this equation; we need to simplify it further. Since we are interested in the large $r$ asymptotic expansion of $g_{iv}^{(1)}$, we can see in the EoM level, the terms on the l.h.s have the order of
\begin{equation}
    \frac{\partial}{\partial r} \left(r^4 f^2(r)\frac{\partial H_{i}^{(1)}}{\partial r}\right) \sim \frac{r^4 f^2(r)H_{i}(r)^{(1)}}{r^2}\, ,
\end{equation}
Then, comparing the order of those two terms at large $r$, we have that
\begin{equation}
    \frac{(1-\frac{1}{Q_{cor}(r)})\frac{4Q^2}{r^2}f(r)H_{i}^{(1)}}{( \frac{\partial}{\partial r}(r^4 f^2(r)\frac{\partial H_{i}^{(1)}}{\partial r})}\Longrightarrow\frac{(1-\frac{1}{Q_{cor}(r)})\frac{4Q^2}{r^2}f(r)}{r^2 f^2(r)}\Big|_{r\to \infty}\sim -\frac{24 r_h^4 (4\pi^2 T^2 +3\frac{T}{C}) }{r^8}+\mathcal{O}(\frac{1}{r^9})\,,
\end{equation}
showing that in the large $r$ limit, the scale of the second term is much more sub-leading than the first term. Therefore, we will ignore the second term in later calculations. The source terms can be written as
\begin{equation}
    s_{i}^{(0)}=f(r)s_{i,2}^{(0)}-\frac{4Q}{r^2 Q_{cor}(r)} \left(c_{i,1}^{(0)}+\int_{r_{h}}^{r}dr' s_{i,1}^{(0)}(r')\right)\, .
\end{equation}
The solution to the quantum-corrected equation \eqref{eq:Diff Eq for H_i with quantum corr} is now given by
\begin{equation}
    H_{i}^{(1)}=-\int_{r}^{\infty}\frac{dr'}{r'^4f(r')^2} \left(c_{i,3}^{(1)}+\int_{r_{h}}^{r'}dr'' s_{i}^{(0)}(r'')\right) + c_{i,4}^{(1)}\, .
\end{equation}
As we can see, the expression is the same as the classical one, and the quantum correction is contained in the source term $s_{i}^{(0)}$. For the $i=y$ case, we can obtain
\begin{equation}
h^{y\, (1)}_{v} = - f(r)\int_{r}^{\infty}\frac{dr'}{r'^4 f(r')^2}\int_{r_{h}}^{r'}dr''s_{y}^{(0)}(r'')\, ,
\end{equation}
where
\begin{equation}
    s_{y}^{(0)} = -\frac{4\sqrt{3}r_{h}^2}{r^2Q_{cor}(r)}\Big(c_{y,1}^{(0)} +i\sqrt{3}r_h^2\omega (\frac{1}{r}-\frac{1}{r_{+}})\delta\beta_{y}\Big)+ 2i\omega \delta\beta_{y} r f(r)\, ,
\end{equation}
and the solution for the gauge field can be expressed as
\begin{equation}
    a_{y}^{(1)} = \frac{1}{4\sqrt{3}r_{h}^2} \left(c_{y,2}^{(1)} + i(r^2-r_{+}^2)\omega\delta \beta_{y} - r^4 \frac{d}{dr} h^{y\, (1)}_{v}\right)\, .
\end{equation}
We present the details of the solutions for $h^{(1)}_{yv}$ and $a^{(1)}_y$ in the large $r$ limit in App.~\ref{App:Pert-Details}.

Similarly, for the $i=x$ case, the only difference is the source term. Then, we can have
\begin{equation}
    h^{x\, (1)}_{v} = -f(r) \int_{r}^{\infty} \frac{dr'}{r'^4f(r')^2 }\int_{r_{h}}^{r'} dr'\, s_{x}^{(0)}(r'')\, ,
\end{equation}
where
\begin{equation}
    s_{x}^{(0)}(r) = 2i\omega rf(r)\delta \beta_x -\frac{4\sqrt{3}r_h^2}{r^2 Q_{cor}(r)} \left(c_{x,1}^{(0)} + i\sqrt{3}r_h (-2k_x \delta r_h + r_h\omega\delta\beta_x)(\frac{1}{r} - \frac{1}{r_{+}})\right)\, .
\end{equation}
We also have
\begin{equation}
    a_{x}^{(1)} = \frac{1}{4\sqrt{3}r_h^2} \left(c_{x,2}^{(1)} + i(r^2 - r_{+}^2)\omega\delta \beta_x - r^4\frac{d}{dr}h_{v}^{x(1)}\right)\, .
\end{equation}
Then, we can obtain the solution at the large $r$ expansion 
\begin{equation}
\begin{aligned}
    h_{yv}^{(1)}\Big|_{r\to \infty} & = \frac{T(2\sqrt{3}c_{y,1} + 3ir_h \omega\delta\beta_y )}{3888 C r_h^4}\Big(224\pi^2 +336\mathrm{arccot}^2(\sqrt{2})-6\pi(144i+39\sqrt{2}-112i\log(r_h))\\
    & \quad + 51\sqrt{2}\arctan (\sqrt{2})\log(3r_h^2)-51\sqrt{2}\mathrm{arccot}(\sqrt{2})\log(6r_h^4)-84(4\arctan^2 (\sqrt{2})\\
    & \quad + \log(2)\log(18)+4\log(r_h)\log(12r_h^4))\Big)r^2 -i\omega \delta\beta_{y}\omega r+\frac{T^2 (2\sqrt{3}c_{y,1}+3ir_h\omega \delta\beta_y)}{1296 C^2 r_h^4}\\
    & \quad \times\Big(2(432 i+117 \sqrt{2}-112\pi)\pi-336\mathrm{arccot^2(\sqrt{2})+336\arctan^2(\sqrt{2})}\\
    & \quad - 51\sqrt{2}\arctan(\sqrt{2})\log(3) + 51\sqrt{2}\mathrm{arccot}(\sqrt{2})\log(6)+84\log(2)\log(18)\\
    & \quad + 6\log(r_h)(-112i\pi + 34\sqrt{2}\mathrm{arccot}(\sqrt{2})-17\sqrt{2}\arctan(\sqrt{2})+56\log(12)\\
    & \quad + 224\log(r_h))\Big) + \frac{1}{1944r_h^3\sqrt{C^5 T}r}\Bigg(2592\sqrt{3}c_{y,1}r_h^4 \sqrt{C^5 T}+2592C^2 r_h^3 T (c_{y,1}\\
    & \quad + 3i\sqrt{3}r_h\omega \delta\beta_y) - 3\sqrt{C T^5} (2\sqrt{3}c_{y,1}+3ir_h\omega\delta\beta_{y})\Big(-18(48i+13\sqrt{2})\pi +224\pi^2 \\
    & \quad + 336\mathrm{arccot}^2 (\sqrt{2}) + 51\sqrt{2}\arctan(\sqrt{2})\log(3) - 51\sqrt{2}\mathrm{arccot}(\sqrt{2})\log(6)\\
    & \quad - 84\log(2)\log(18)+6\log(r_h)(112i\pi -34 \sqrt{2}\mathrm{arccot}(\sqrt{2})+17\sqrt{2}\arctan(\sqrt{2})\\
    & \quad - 56\log(12)-224\log(r_h))\Big)+2r_h^2(CT)^{3/2}(-2i\sqrt{3}c_{y,1} + 3r_h \omega\delta\beta_y)\Big(672\pi\log(r_h)\\
    & \quad + 6i\log(r_h)(-648 + 34\sqrt{2}\mathrm{arccot}(\sqrt{2}) - 17\sqrt{2}\arctan(\sqrt{2})+56\log(12)\\
    & \quad + 224\log(r_h)) + i(1944 + 2 (432i + 117\sqrt{2} - 112\pi) \pi - 336 \mathrm{arccot}^2 (\sqrt{2})\\
    & \quad + 336 \arctan^2(\sqrt{2}) - 51\sqrt{2}\arctan(\sqrt{2})\log(3) + 51\sqrt{2}\mathrm{arccot}(\sqrt{2})\log(6)\\
    & \quad + 84 \log(2)\log(18) + 1944\log(\frac{3T}{C}))\Big)\Bigg) + \mathcal{O}(\frac{1}{r^2})\,.
\end{aligned}
\end{equation}
And 
\begin{equation}
\begin{aligned}
      a_{y}^{(1)} \Big|_{r\to \infty} & = \frac{T^2 (2c_{y,1}+i\sqrt{3}r_h \omega \delta \beta_y)r}{2592 C^2 r_h^6} \Big(2(432 i +117 \sqrt{2}-112\pi)\pi -336\mathrm{arccot}^2 (\sqrt{2})\\
      & \quad - 51\sqrt{2}\arctan(\sqrt{2})\log(3) + 51\sqrt{2}\mathrm{arccot}(\sqrt{2})\log(6)+84\log(2)\log(18)\\
      & \quad + 6\log(r_h)(-112i\pi + 34\sqrt{2}\mathrm{arccot}(\sqrt{2})-17\sqrt{2}\arctan(\sqrt{2})+56\log(12)\\
      & \quad + 224\log(r_h))\Big) + \frac{1}{2592r_h^5 \sqrt{C^5 T}}\Bigg(216\sqrt{3}c_{y,2}r_h^3 \sqrt{C^5 T}-864 iC^3 \pi^2 r_h^4 T^2 \omega \delta \beta_y\\
      & \quad + 432C^2 r_h^3 T (2\sqrt{3}c_{y,1} + i(15r_h -2\sqrt{3}\pi^2 T\sqrt{CT})\omega \delta \beta_y)-3i\sqrt{3}r_h \omega \delta\beta_{y}\Big(72 r_h^4 \sqrt{C^5 T}\\
      & \quad + \sqrt{C T^5}(-18(48i + 13\sqrt{2})\pi +224\pi^2 +336 \mathrm{arccot}^2(\sqrt{2})-336\arctan^2(\sqrt{2})\\
      & \quad + 51\sqrt{2}\arctan(\sqrt{2})\log(3) - 51\sqrt{2}\mathrm{arccot}(\sqrt{2})\log(6)-84\log(2)\log(18))\Big)\\
      & \quad + 6c_{y,1}\Big(432r_h^4 \sqrt{C^5 T}+\sqrt{C T^5}(2(432i+117\sqrt{2}-112\pi)\pi -336\mathrm{arccot}^2(\sqrt{2})\\
      & \quad + 336\arctan^2(\sqrt{2}) - 51\sqrt{2}\arctan(\sqrt{2})\log(3) + 51\sqrt{2}\mathrm{arccot}(\sqrt{2})\log(6)\\
      & \quad + 84\log(2)\log(18)\Big) + 2r_h^2 (CT)^{3/2}\Big(i\sqrt{3}r_h \omega\delta\beta_y(1620+2(432i+117 \sqrt{2}\\
      & \quad - 112\pi)\pi-336\mathrm{arccot}^2(\sqrt{2}) + 336\arctan^2(\sqrt{2})-51\sqrt{2}\arctan(\sqrt{3})\log(3)\\
      & \quad + 51\sqrt{2}\mathrm{arccot}(\sqrt{2})\log(6) + 84\log(2)\log(18))+2c_{y,1}(1944+2(432i+117\sqrt{2}\\
      & \quad - 112\pi)\pi - 336 \mathrm{arccot}^2(\sqrt{2}) + 336\arctan^2(\sqrt{2})-51\sqrt{2}\arctan(\sqrt{2})\log(3)\\
      & \quad + 51\sqrt{2}\mathrm{arccot}(\sqrt{2})\log(6)+84\log(2)\log(18))\Big)+6(2c_{y,1}+i\sqrt{3}r_h \omega \delta \beta_y)\\
      & \quad \times\Big((2r_h^2 (CT)^{3/2}(-648-112i\pi +34\sqrt{2}\mathrm{arccot}(\sqrt{2})-17\sqrt{2}\arctan(
      \sqrt{2})\\
      & \quad + 112\log(2)+56\log(3))+3\sqrt{C T^5}(-112i\pi +34\sqrt{2}\mathrm{arccot}(\sqrt{2})\\
      & \quad - 17\sqrt{2}\arctan(\sqrt{2}) + 56\log(12))\Big)\log(r_h)+224(2r_h^2 (CT)^{3/2}\\
      & \quad + 3\sqrt{CT^5})\log(r_h)^2 + 648r_h^2 (CT)^{3/2}\log(\frac{3T}{C})\Bigg)+\frac{1}{1296 r_h^4 \sqrt{C^5 T}r}\\
      & \quad \times \Bigg(-3888iC^2 r_h^4 T \omega \delta\beta_y+r_h^2 (CT)^{3/2}\Big(3888i\sqrt{3}r_h \omega\delta\beta_y +(2c_{y,1}+i\sqrt{3}r_h \omega\delta \beta_y)\\
      & \quad \times (-18(48i +13\sqrt{2})\pi +224\pi^2 +336\mathrm{arccot}^2(\sqrt{2})-336\arctan^2(\sqrt{2})\\
& \quad + 51\sqrt{2}\arctan(\sqrt{2})\log(3) - 51\sqrt{2}\mathrm{arccot}(\sqrt{2})\log(6)-84\log(2)\log(18)
)\Big)\\
& \quad -3(2c_{y,1}(216r_h^4 \sqrt{C^5 T} + \sqrt{C T^5}(2\pi(-432 i -117 \sqrt{2}+112\pi)+336\mathrm{arccot}^2 (\sqrt{2})\\
& \quad -336\arctan^2 (\sqrt{2}) + 51\sqrt{2}\arctan(\sqrt{2})\log(3)-51\sqrt{2}\mathrm{arccot}(\sqrt{2})\log(6)\\
& \quad -84\log(2)\log(18))) + \sqrt{3}r_h\omega\delta\beta_y(-432ir_h^4 \sqrt{C^5 T}+\sqrt{C T^5}(2(432-117i\sqrt{2}\\
& \quad + 112i\pi)\pi + 3i(122\mathrm{arccot}^2(\sqrt{2})-112\arctan^2(\sqrt{2})+17\sqrt{2}\arctan(\sqrt{2})\log(3)\\
& \quad - 17\sqrt{2}\mathrm{arccot}(\sqrt{2})\log(6)-28\log(2)\log(18))))+6(-r_h^2 (CT)^{3/2}+3\sqrt{CT^5})\\
& \quad \times (2c_{y,1}+i\sqrt{3}r_h\omega\delta\beta_y)\log(r_h)(-112i\pi +34 \sqrt{2}\mathrm{arccot}(\sqrt{2})-17\sqrt{2}\arctan(\sqrt{2})\\
& \quad + 56\log(12)+224\log(r_h))\Bigg)+\mathcal{O}(\frac{1}{r^2})\,.
\end{aligned}
\end{equation}

The solutions for $h^{(1)}_{xv}$ and $a^{(1)}_x$ are:

\begin{equation}
    \begin{aligned}
        h_{xv}^{(1)}\Big|_{r\to \infty} & = -i\omega \delta\beta_{x}r +\Bigg(\frac{4c_{x,1}r_h}{\sqrt{3}}+4ik_x r_h\delta r_h+\frac{4}{3}\sqrt{\frac{T}{C}}(c_{x,1}-3i\sqrt{3}(2k_x\delta r_h-r_h\delta\beta_{x}\omega))\\
        & \quad + \frac{2T(2\sqrt{3}c_{x,1}-6ik_x \delta r_h+3ir_h \omega \delta \beta_x)(1+\log(\frac{3T}{C r_h^2}))}{C r_h}\Bigg) + \mathcal{O}(\frac{1}{r^2})\,,
    \end{aligned}
\end{equation}
and
\begin{equation}
    \begin{aligned}
        a_{x}^{(1)}\Big|_{r\to \infty} & = \frac{1}{12 r_h^3 C}\Big(4c_{x,1}(9T+\sqrt{3}r_h\sqrt{CT}) + 3i(\sqrt{3}T+2r_h\sqrt{CT})(-12k_x \delta r_h +5r_h \omega \delta\beta_{x})\\
        & \quad + Cr_h(\sqrt{3}c_{x,2}+12c_{x,1}r_h -i(4\pi^2 r_h T\sqrt{CT}\omega\delta\beta_x+\sqrt{3}(-12k_x r_h\delta r_h +(r_h^2 \\
        & \quad + 4\pi^2 T^2)\omega\delta\beta_x))) +18T(2c_{x,1}-i\sqrt{3}(2k_x\delta r_h -r_h \omega\delta\beta_x))\log(\frac{3T}{Cr_h^2})\Big)\\
        & \quad + \frac{1}{Cr_h^2 r}(-c_{x,1}Cr_h^2 +i(\sqrt{3}Cr_h^2 +3\sqrt{3}T-3r_h \sqrt{CT})(-2k_x\delta r_h+r_h\omega\delta\beta_{x}))\\
        & \quad + \mathcal{O}(\frac{1}{r^2})\, .
    \end{aligned}
\end{equation}

\subsubsection{Scalar Sector}

In the scalar sector, the relevant components are $g_{vv}^{(n)}$, $g_{vr}^{(n)}$, and $A_{v}^{(n)}$, while $g_{xx}^{(n)} + g_{yy}^{(n)}$ can be determined from $g_{vr}^{(n)}$ by using the gauge condition. Consequently, we have
\begin{equation}
    h_{xx}^{(1)} + h_{yy}^{(1)} = 2 r^2 \sigma^{(1)}\, ,\quad h_{vr}^{(1)} = -\sigma^{(1)}(r)\, .
\end{equation}
The classical EoM for $h_{vr}$ is
\begin{equation}
    \frac{1}{r} \Big[2h_{vr}'(r) - 2\sigma'(r) - r\sigma''(r)\Big] + \frac{ik_{x}}{r^2} \Big[r^2 {h^{x}}_{r} (r)\Big]' = 0\, .
\end{equation}
Then, the EoM at the $n$-th order has the following general expression:
\begin{equation}
    \frac{\partial}{\partial r} \left(r^4 \frac{\partial}{\partial r} g_{vr}^{(n)}\right) = s_{vr}^{(n)}\, ,
\end{equation}
whose general solution is
\begin{equation}
    g_{vr}^{(n)}=-\int_{r}^{\infty}\frac{dr'}{r'^4}\int_{r_{B}}^{r'}dr'' s_{vr}^{(n)}(r'') +\frac{c_{vr,1}^{(n)}}{r^3}+c_{vr,2}^{(n)}\, .
\end{equation}
For the $n=1$ case, the source term equals 0, 

\begin{equation}
    s_{vr}^{(1)} = -ik_{x}r^2 \frac{d}{dr}h_{xr}^{(0)} = 0\, .
\end{equation}

Since there is no metric factor $f(r)$, the quantum-corrected EoM is the same as the classical one. Hence, we have
\begin{equation}
    \left(r^4 \frac{d}{dr} h_{vr}^{(1)}(r)\right)' = 0\, ,
\end{equation}
whose solution can be written as
\begin{equation}
    h_{vr}^{(1)} = \frac{c_{vr,1}^{(1)}}{r^3}\, .
\end{equation}
Using the residual gauge freedom of gauge condition eq.(\ref{Gauge choice}), we can set $c_{vr,1}^{(1)} = 0$ \cite{Moitra:2020dal}, and then obtain $\sigma^{(1)} =-h_{vr}^{(1)} =0$. Similarly, The EoM for $a_{v}^{(1)}$ is
\begin{equation}
    \frac{d}{dr}(r^2 \frac{d}{dr}A_{v}^{(n)})=\hat{s}^{(n)}_{v}\, ,
\end{equation}
and the source term equals
\begin{equation}
    \hat{s}_{v}^{(0)}=-r^2 s_{v}^{(0)}+2Q\frac{d}{dr}h_{vr}^{(0)}=-\frac{6\sqrt{3}c_{vr,1}r_h^2}{r^4}=0\, ,
\end{equation}
where
\begin{equation}
    s^{(0)}_{v} = i k_x \left(\frac{d g(r)}{dr} r^{-2}h_{xr}^{(0)}-\frac{1}{r^2}\frac{d a_{x}^{(0)}}{dr}\right) = 0\, .
\end{equation}
Therefore, we have
\begin{equation}
    a_{v}^{(1)} = \frac{c_{v,1}^{(1)}}{r}\, .
\end{equation}
The EoM for $g_{vv}^{(n)}$ is 
\begin{equation}
    \frac{d}{dr}(r^2\frac{d}{dr}g_{vv}^{(n)})=s_{vv}^{(n)}\, ,
\end{equation}
where the source term is 
\begin{equation}
    \begin{aligned}
        s_{vv}^{(0)} & = \frac{d}{dr}(r^4f(r)Q_{cor}(r)\frac{d}{dr}\sigma^{(1)}(r))-\frac{d}{dr}(r^4 f(r)Q_{cor}(r))\frac{d}{dr}h_{vr}^{(1)}-12r^2h_{vr}^{(1)}-4r^2\frac{d}{dr}(g(r))\frac{d}{dr}a_{v}^{(1)}\\
        & \quad - i\Big(k_x \frac{d}{dr}(r^2 f(r)Q_{cor}(r)h_{xr}^{(0)}+h_{xv}^{(0)})+2\omega r^2 \frac{d}{dr}h_{vr}^{(1)}+2\omega r\frac{d}{dr}(r\sigma^{(1)}(r))\Big)\\
        & = \frac{2}{Cr^3}(6ik_x r_h (r+3r_h)T\delta\beta_x+8iCk_x\pi^2 r_h(r+3r_h)T^2 \delta\beta_x +Cr(2\sqrt{3}c_{v,1}r_h^2+ik_x r^3 \delta\beta_x)\, ,
    \end{aligned}
\end{equation}
which leads to the solution
\begin{equation}
\begin{aligned}
    h_{vv}^{(1)} & = \frac{1}{3C^3 r_h^4 r^3}\Bigg(-4C^{7/2}\pi^2 r^2 (r-r_h)r_h^3 T^{3/2} (6c_{v,1}+i\sqrt{3}k_x r_h \delta\beta_x)\\
    & \quad - 6r^2(r-r_h)r_h^3 \sqrt{C^5 T}(6c_{v,1}+i\sqrt{3}k_x r_h \delta\beta_x)+9C^2 r_h^2 T(2\sqrt{3}c_{v,1}r^2 (3r-2r_h)\\
    & \quad + ik_x r_h (6r^3-9r^2 r_h +2rr_h^2 +2r_h^3)\delta\beta_x)+18Cr^2 T (3c_{v,1}(\sqrt{3}(5r-2r_h)T \\ & \quad - 4rr_h\sqrt{CT}+2r_h^2 \sqrt{CT} )+ik_x r_h(45rT-33r_hT -8\sqrt{3}rr_h \sqrt{CT}\\
    & \quad + 8\sqrt{3}r_h^2 \sqrt{CT})\delta\beta_x)+3C^3 r_h^2 (4\pi^2 T^2(2\sqrt{3}c_{v,1}r^2(3r-2r_h)+ik_xr_h(6r^3 \\
    & \quad - 9r^2 r_h +2rr_h^2 +2r_h^3)\delta\beta_x)+rr_h^2 (2\sqrt{3}c_{v,1}(r-r_h)^2+r(c_{vv,1}+c_{vv,2}r\\
    & \quad + ik_x (r-r_h)^2 \delta\beta_{x})\Bigg)\, .    
\end{aligned}
\end{equation}
From the gauge fixing condition eq.(\ref{Gauge choice}), we have
\begin{equation}
\begin{aligned}
   h_{vr}^{(1)}&=-\sigma^{(1)}=0\\
   h_{rr}^{(0)}&=h_{rx}^{(1)}=h_{ry}^{(1)}=a_{r}^{(1)}=0\,.
\end{aligned}
\end{equation}
Now that we have obtained all the first-order metric and gauge field component perturbation solutions, some of the integration constants remain unfixed here. The Landau frame condition allows us to fix them (See App.~\ref{sec: landau frame condition} for details).

\subsection{Hydrodynamic Modes}

 In this subsection, we will show the details of using the first-order constitutive relation and the conservation equations of boundary fluid to calculate the dispersion relation.
The conservation equations $\partial_\mu T^{\mu\nu}=0$ and $\partial_\mu J^\mu=0$ lead to the dispersion relation for the hydrodynamic modes.

\begin{equation}
    \begin{aligned}
        3k_x r_h (3T +2 C (r_h^2 +2 \pi^2 T^2))\delta\beta_x-2((6Cr_h^2 +3T +4C \pi^2 T^2)\delta r_h+r_h (3+8C \pi ^2 T^2)\delta T)\omega&=0\,,\\
        6\sqrt{3}C^{3/2}k_x^2 r_h^3 \sqrt{T}\delta\beta_x+144\sqrt{3}\sqrt{C}k_x^2 r_h T^{3/2}\delta\beta_x -594k_x^2 T^2 \delta\beta_x +18iCr_h^2 (k_x r_h \delta T +k_x T(\delta r_h &\\
        +5ik_x\delta \beta_x )-3r_h T \omega \delta\beta_x)+3C^2r_h^2 (k_x (16i \pi r_h T \delta T +8i\pi^2 T^2(\delta r_h +5ik_x \delta\beta_x)+r_h^2(12i\delta r_h &\\+k_x \delta\beta_x))-12ir_h (r_h^2 +2\pi^2 T^2)\omega\delta\beta_x)  &=0\,,\\
        4\sqrt{3} C^{5/2}k_x^2 \pi^2 r_h^3 T^{3/2}+6k_x^2 (-99T^2 +\sqrt{3}r_h(r_h^2 \sqrt{C^3 T}+24\sqrt{C T^3}))-18C r_h^2 T(5k_x^2 +3 ir_h \omega)&\\
        +3C^2 r_h^2 (k_x^2 (r_h^2 -40 \pi^2 T^2)-12i r_h (r_h^2 +2\pi^2 T^2)\omega)&=0\,,\\
         i\sqrt{3}k_x r_h^2 \delta\beta_x -2i\sqrt{3}r_h \omega \delta r_h - k_x \frac{(\sqrt{3}C r_h^2 +3\sqrt{3}T-3r_h \sqrt{C T})(-2k_x \delta r_h +r_h \omega \delta\beta_x)}{C r_h^2}&\\
        + \Big(6C k_x r_h^2 \delta r_h +3(3T +2\sqrt{3}r_h \sqrt{C T})(-2k_x \delta r_h+r_h \omega \delta\beta_x)+9T(-2k_x \delta r_h &\\+r_h \omega \delta\beta_x)\log(\frac{3T}{C r_h^2})\Big)/ \left(2\sqrt{3}C r_h^2 +2r_h \sqrt{CT}+6\sqrt{3}T (1+\log(\frac{3T}{C r_h^2})) \right) & = 0\,.
    \end{aligned}
\end{equation}
Where  the first 3 equations are from $\partial_{\mu}T^{\mu \nu}=0$, and the last one is from $\partial_{\mu}J^{\mu}=0$. This  four equations are homogeneous equations for $\delta T, \delta r_{h}, \delta \beta_{x},\delta\beta_{y} $, and the coefficients contains $\omega,k_{x}$. These equations have a non-trivial solution if and only if the determinant of the coefficients equals zero \cite{Bhattacharyya:2007vjd}. From the determinant of the coefficient equal to zero, we can also obtain four equations for four hydrodynamic modes, the first one is

\begin{equation}
    \begin{aligned}
        &6\sqrt{3}C^{3/2}k_x^2 r_h^3\sqrt{T}+144\sqrt{3}\sqrt{C}k_x^2 r_hT^{3/2} +4\sqrt{3}C^{5/2}k_x^2 \pi^2 r_h^3 T^{3/2}-594k_x^2 T^2\\
        &18Cr_h^2 T(5k_x^2 +3ir_h \omega)+3C^2 r_h^2 (k_x^2(r_h^2 -40\pi^2 T^2 )-12ir_h (r_h^2 +2\pi^2 T^2)\omega)=0\,.
    \end{aligned}
\end{equation}

From this, we can obtain the shear mode as

\begin{equation}
\begin{aligned}
        \omega & = -\frac{ik_x^2}{18C r_h^3 (3T +2 C (r_h^2 +2\pi^2 T^2))}(-90Cr_h^2 T +4\sqrt{3}C^{5/2}\pi^2 r_h^3 T^{3/2}-594 T^2 \\ & \quad + 3C^2(r_h^4 -40\pi^2 r_h^2 T^2)+6\sqrt{3}r_h(r_h^2\sqrt{C^3 T}+24\sqrt{C T^3}))\,.
\end{aligned}
\end{equation}

The second one is

\begin{equation}
\begin{aligned}
     &-6i\sqrt{3}C^{3/2}k_x^2 r_h63 \sqrt{T}\omega -144 i \sqrt{3}\sqrt{C}k_x^2 r_h T^{3/2}\omega -4i \sqrt{3}C^{5/2}k_x^2 \pi^2 r_h^3 T^{3/2}\omega\\
     &+594 ik_x^2 T^2 \omega +9Cr_h62 T(k_x^2 (3r_h +10 i \omega)-6 r_h \omega^2)+3C^2 r_h^2 (-12 r_h(r_h62 +2\pi^2 T^2)\omega^2 \\
     &+k_x^2(6r_h^3 +12\pi^2 r_h T^2 -ir_h^2 \omega +40 i \pi^2 \omega T^2 ))=0\,,
\end{aligned}
\end{equation}

From which we can solve the two sound modes as

\begin{equation}
    \begin{aligned}
        \omega &=\pm\frac{k_x}{\sqrt{2}}-\frac{ik_x^2}{288 C^2 r_h^5}(12 C^2 r_h^4 +540 \sqrt{3}\sqrt{C}r_h T^{3/2}-1809 T^2 
        +8\sqrt{3}C^{3/2}r_h^3 \sqrt{T}(3+2C\pi^2 T)
        \\ & \quad - 126 C r_h^2 T (3+4C \pi^2 T))+\mathcal{O}(k_x^3,T^{5/2})\,.
    \end{aligned}
\end{equation}

Finally, we have
\begin{equation}
    \begin{aligned}
        &3C k_x^2 r_h^2 (29 T +\sqrt{3}r_h \sqrt{CT})-k_x^2 T (25T +74\sqrt{3}r_h \sqrt{CT})+9C^2 r_h^4(k_x^2 -2ir_h \omega)\\
        &-9k_x^2 T(-6C r_h^2 +20T +9\sqrt{3}r_h \sqrt{C T})\log (\frac{3T}{C r_h^2})-162 k_x^2 T^2 \left(\log(\frac{3T}{C r_h^2})\right)^2 = 0\,.
    \end{aligned}
\end{equation}
Solving this, we can obtain the charge diffusion mode as
\begin{equation}
\begin{aligned}
        \omega &=-\frac{ik_x^2}{18 C^2 r_h^5}\Big( 9C^2 r_h^4 +3 C r_h^2 (29T +\sqrt{3}r_h \sqrt{CT})-T(25 T +74 \sqrt{3}r_h \sqrt{CT})\\
    & \quad - 9T \log(\frac{3T}{C r_h^2})(-6Cr_h ^2 +20 T +9\sqrt{3}r_h \sqrt{C T}+18 T \log(\frac{3T}{C r_h^2}))\Big)\,.
\end{aligned}
\end{equation}
 The final answer has been rewritten in a standard form in Sec.~\ref{sec:Near-Ext Fluid Gravity Correspondence}.

\section{Landau Frame Condition}\label{sec: landau frame condition}
In this section, we will determine the integration constant that appears in the first-order calculation by using the Landau frame condition.

The Landau frame condition, also called the Landau-Lifshitz frame condition\cite{landau2013fluid, Moitra:2020dal}, requires
\begin{equation}
    u^{\mu}T_{\mu\nu}^{(n)}=0,\quad u^{\nu}J_{\nu}^{(n)}=0\,,
\end{equation}
which fixes the residual gauge freedom of the metric and gauge field. The physical meaning of the Landau frame condition is that we restrict the transportation direction of boundary fluid velocity to be the same transportation direction as the gravity solution. This requirement explicitly makes 
\begin{equation}
    T_{vv}^{(1)}=0,\quad T_{vx}^{(1)}=0,\quad T_{vy}^{(1)}=0,\quad J_{v}^{(1)}=0\, .
\end{equation}
Where $T_{vv}^{(1)}=0$ makes
\begin{equation}
\begin{aligned}
      &c_{vv,1}^{(1)}-\frac{4\sqrt{3}c_{v,1}^{(1)}(r_h^2 +4\pi^2 T^2)}{r_h}  +ik_x (r_h -6\pi T)(r_h+6\pi T)\delta\beta_{x}-\frac{3T(4\sqrt{3}c_{v,1}^{(1)}+9ik_x r_h \delta \beta_x)}{C r_h}\\
      &+\frac{4}{3}\sqrt{C\pi^2 }T^{3/2}(6c_{v,1}^{(1)}+i\sqrt{3}k_x r_h \delta\beta_x)+\frac{2T (6c_{v,1}^{(1)}+i\sqrt{3}k_x r_h \delta\beta_x)}{\sqrt{CT}}+\frac{6T}{C^2r_h^3}(c_{v,1}^{(1)}(-6\sqrt{3}T\\
      &+6r_h\sqrt{CT})+ik_x r_h(-33 T+8\sqrt{3}r_h \sqrt{CT})\delta\beta_{x})=0\,.
\end{aligned}
\end{equation}
Then we can fix $c_{vv,1}$ in terms of $c_{v,1}$. For the $T_{vx}^{(1)}$ component, we have
\begin{equation}
    \begin{aligned}
        &\frac{1}{3C r_h} \Big(4c^{(1)}_{x,1}(3\sqrt{3}T+r_h \sqrt{CT})+4Cr_h^2 (\sqrt{3}c^{(1)}_{x,1}+3ik_x \delta r_h)+6i(3T+2\sqrt{3}r_h\sqrt{CT})\\
        &\times(-2k_x \delta r_h+r_h \omega\delta \beta_x)+6T(2\sqrt{3}c^{(1)}_{x,1}-6ik_x \delta r_h+3i r_h \omega \delta\beta_x)\log(\frac{3T}{C r_h^2}) \Big) = 0\,. 
    \end{aligned}
\end{equation}
Which fixes the $c_{x,1}^{(1)}$ as
\begin{equation}
    \begin{aligned}
        c_{x,1}^{(1)}&=\frac{3i}{2\sqrt{3}Cr_h^2 +2r_h \sqrt{C T}+6\sqrt{3}T(1+\log(\frac{3T}{C r_h^2}))} \Big(-2C k_x r_h^2 \delta r_h +(3T+2\sqrt{3}r_h \sqrt{C T})\\
        & \quad \times(2k_x \delta r_h -r_h \omega \delta \beta_x)+(6k_x T \delta r_h -3r_h T \omega \delta\beta_x) \log(\frac{3T}{C r_h^2})\Big)\,.
    \end{aligned}
\end{equation}
For the $T_{vy}^{(1)}$ components, we have
\begin{equation}
    \begin{aligned}
        &\frac{1}{1944r_h^3\sqrt{C^5 T}}\Bigg(2592\sqrt{3}c_{y,1}r_h^4 \sqrt{C^5 T}+2592C^2 r_h^3 T (c_{y,1}\\
    &+3i\sqrt{3}r_h\omega \delta\beta_y)-3\sqrt{C T^5} (2\sqrt{3}c_{y,1}+3ir_h\omega\delta\beta_{y})\Big(-18(48i+13\sqrt{2})\pi +224\pi^2 \\
    &+336\mathrm{arccot}^2(\sqrt{2})+51\sqrt{2}\arctan(\sqrt{2})\log(3)-51\sqrt{2}\mathrm{arccot}(\sqrt{2})\log(6)\\
    &-84\log(2)\log(18)+6\log(r_h)(112i\pi -34 \sqrt{2}\mathrm{arccot}(\sqrt{2})+17\sqrt{2}\arctan(\sqrt{2})\\
    &-56\log(12)-224\log(r_h))\Big)+2r_h^2(CT)^{3/2}(-2i\sqrt{3}c_{y,1}+3r_h \omega\delta\beta_y)\Big(672\pi\log(r_h)\\
    &+6i\log(r_h)(-648+34\sqrt{2}\mathrm{arccot}(\sqrt{2})-17\sqrt{2}\arctan(\sqrt{2})+56\log(12)\\
    &+224\log(r_h))+i(1944+2(432i+117\sqrt{2}-112\pi)\pi-336\mathrm{arccot}^2(\sqrt{2})\\
    &+336\arctan^2(\sqrt{2})-51\sqrt{2}\arctan(\sqrt{2})\log(3)+51\sqrt{2}\mathrm{arccot}(\sqrt{2})\log(6)\\
    &+84\log(2)\log(18)+1944\log(\frac{3T}{C}))\Big)\Bigg)=0\,.
    \end{aligned}
\end{equation}
Which can fixes $c_{y,1}$. As for $J_{\nu}^{(1)}$, we have
\begin{equation}
    c_{v,1}^{(1)}=0\,.
\end{equation}
Therefore, by choosing the Landau frame condition, all residual integration constants in the main text have been fixed.

\bibliographystyle{JHEP}
\bibliography{refs}
\end{document}